\documentclass[aps,prx,twocolumn,superscriptaddress,longbibliography]{revtex4-2}

\usepackage{subcaption}
\usepackage{graphicx}
\usepackage{amsmath}
\usepackage{amssymb}
\usepackage{hyperref}
\usepackage{xcolor}
\usepackage{siunitx}

\newcommand{\fb}[1]{\textcolor{red}{FB: #1}}

\DeclareMathOperator*{\argmin}{arg\,min}

\begin{document}

% Title
\title{Memristive Linear Algebra}

% Authors and affiliations
\author{J. Lin}
\email{jlin1212@lanl.gov}
\affiliation{Theoretical Division (T-4), Los Alamos National Laboratory, New Mexico, 87545, USA}
\affiliation{Center for Nonlinear Studies, Los Alamos National Laboratory, New Mexico, 87545, USA}

\author{F. Barrows}
\email{fbarrows@lanl.gov}
\affiliation{Theoretical Division (T-4), Los Alamos National Laboratory, New Mexico, 87545, USA}
\affiliation{Center for Nonlinear Studies, Los Alamos National Laboratory, New Mexico, 87545, USA}

\author{F. Caravelli}
\email{caravelli@lanl.gov}
\affiliation{Theoretical Division (T-4), Los Alamos National Laboratory, New Mexico, 87545, USA}
% Abstract
\begin{abstract}
The advent of memristive devices offers a promising avenue for efficient and scalable analog computing, particularly for linear algebra operations essential in various scientific and engineering applications. This paper investigates the potential of memristive crossbars in implementing matrix inversion algorithms. We explore both static and dynamic approaches, emphasizing the advantages of analog and in-memory computing for matrix operations beyond multiplication. Our results demonstrate that memristive arrays can significantly reduce computational complexity and power consumption compared to traditional digital methods for certain matrix tasks. Furthermore, we address the challenges of device variability, precision, and scalability, providing insights into the practical implementation of these algorithms.
\end{abstract}

% Make title
\maketitle

% Introduction
%\tableofcontents
%\clearpage

\begin{figure*}
    \centering
    \includegraphics[width=\linewidth]{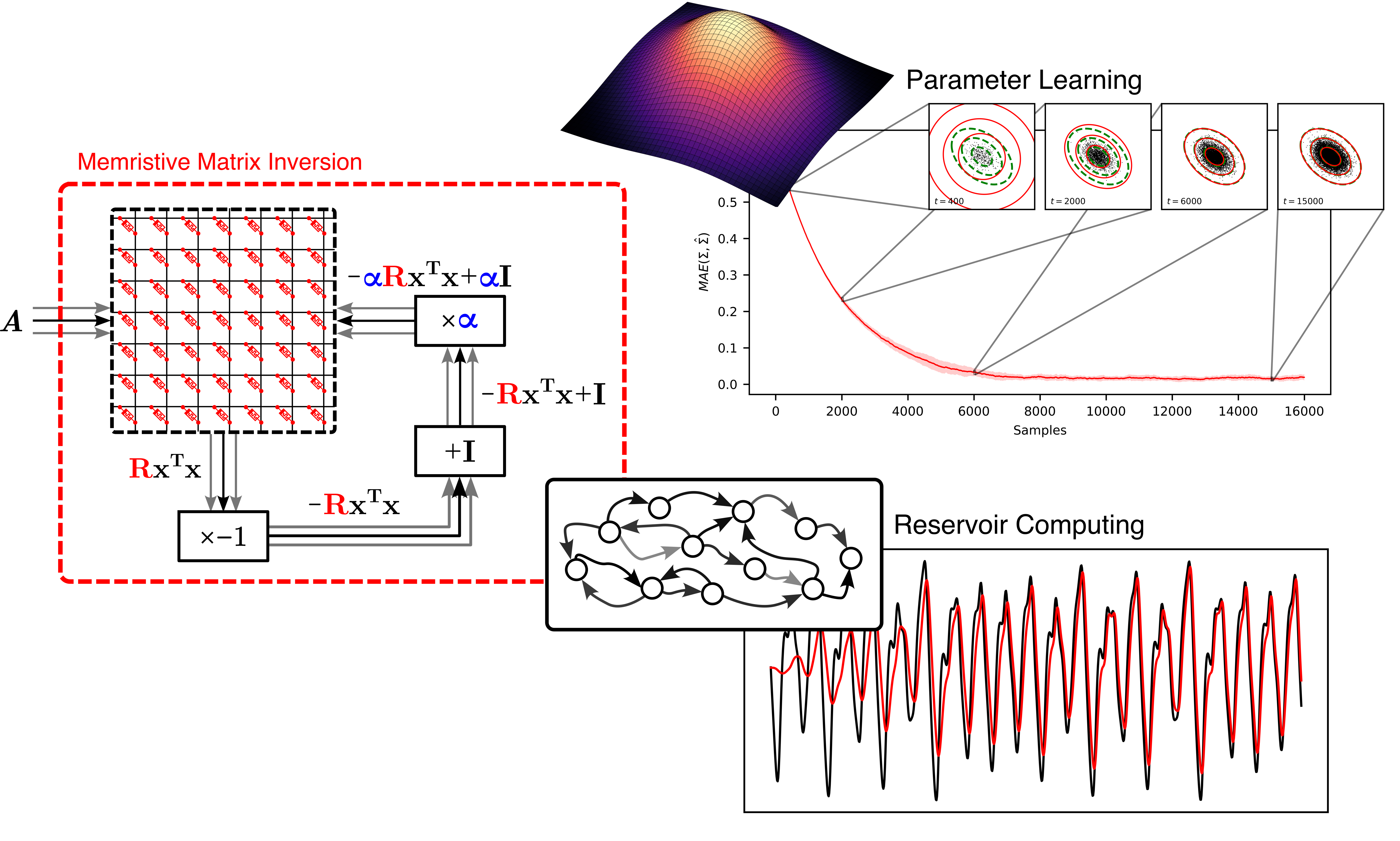}
    \caption{The matrix inversion algorithm leverages Kirchhoff's laws to compute a feedback-based matrix inversion with computational advantage. At each iteration, a matrix $\mathbf{A}$ is right-multiplied with the current crossbar state in $O(N)$ and used to compute an incremental update to the crossbar state until the crossbar state $\mathbf{R}^* = \mathbf{A}^{-1}$. Aside from the matrix-matrix multiplication, only three applications of simple arithmetic are required. The resulting process is simple and powerful, and power consumption and speed can be tuned with the control parameter $\alpha$. This enables the implementation of output filters for applications like parameter learning and reservoir computing.}
    \label{fig:overview}
\end{figure*}

\section{Introduction}

Matrix and matrix-vector operations are fundamental operations in various scientific and engineering applications, including solving systems of linear equations \cite{golub13}, signal processing \cite{Ryan2019LinearAS}, scientific computing \cite{Heath2018,Teukolski}, machine learning \cite{aggarwal}, and control systems \cite{Datta1994}. Various algorithms are available for inverting general matrices, such as Gaussian elimination, Gauss-Jordan elimination \cite{Althoen1987}, Cholesky decomposition \cite{Menon}, QR decomposition, and LU decomposition \cite{Teukolski}. These algorithms, while common, are computationally demanding and typically involve cubic complexity in terms of the number of matrix-vector operations. State-of-the-art algorithms for matrix inversion will be reviewed in Sec. \ref{sec:sota}.

These methods, while well-established, face significant challenges related to memory consumption, algorithmic complexity, and power efficiency, particularly as the size of the matrices increases. High-performance GPUs are known for their significant power consumption (300-350 W), which is a critical factor in designing energy-efficient computing systems \cite{wang2021characterization} \footnote{For instance, the NVIDIA GeForce RTX 3090 has an estimated power consumption of approximately 350 watts (W), while the NVIDIA GeForce RTX 3080 consumes around 320 W. Similarly, the AMD Radeon RX 6900 XT and AMD Radeon RX 6800 XT both have estimated power consumptions of approximately 300 W each.}. 

In recent years, there has been growing interest in leveraging memristive crossbar arrays for computational tasks due to their potential for high-density integration, non-volatility, and low power consumption \cite{yu2018}. Memristive devices (e.g. resistors with memory) are resistive switching devices that can be used to perform analog matrix-vector multiplications efficiently. These are promising candidates for implementing analog, neuromorphic, and other unconventional computing paradigms.

Over the last few years, using the fact that Kirchhoff laws can be exploited to perform matrix-vector multiplication operations in one-shot,  memristive crossbars have been shown to be a promising platform to implement matrix operation algorithms. 

However, despite these promising developments, several challenges remain in realizing practical memristive crossbar-based matrix inversion. One major issue is the precision of the analog computations, which can be affected by device variability, non-idealities, and noise. Despite these challenges, research in crossbar arrays has flourished over the last decade \cite{Alibart2013,Xia_2009,Jo_2010,Pi2018,Christensen2022,Xia2019,WanGeeKim2014,Ohba2018,Woo2018,Wang2018}, and state-of-the-art cross-point memristive memory has reached effectively 1024 (10 bits) states using a variety of noise-reduction techniques \cite{Song_2024}. Although this precision is not enough yet for scientific computing, the methods currently implemented for noise reduction are scalable and promising for error mitigation in future technology. 

Crossbars are typically thought of as accelerators of matrix-matrix and matrix-vector multiplication \cite{mehonic}. However, \cite{Sun2019} present an innovative approach for matrix inversion utilizing cross-point resistive arrays. The process involves implementing the target matrix in a cross-point array circuit, where the input currents are applied, and the resulting output potentials are measured. This physical realization leverages the properties of resistive memory devices (RRAM) to perform matrix-vector multiplication (MVM) efficiently. The matrix inversion is achieved through a feedback mechanism with operational amplifiers that forces the output voltage to satisfy the equation $A \cdot V + I = 0$, thus obtaining $V = -A^{-1} \cdot I$. In practice, the authors solve linear equations $A \vec x=\vec b$ in one step, but this can be used to solve for matrix inverses in $N$ steps by carefully choosing $\vec b$ at each iteration, where $N$ is the size of the matrix. The authors also demonstrated this method's accuracy and stability by comparing the experimentally measured inverse matrix with small matrices. 

We ask whether it is possible to use a different method to solve a variety of problems at the same time. It is also crucial to note that the method in \cite{Sun2019}, while very fast, requires a cross-point array whose junction conductances are already driven to represent $A$. Our method addresses this requirement by instead driving the physical state of the cross-point array to $A^{-1}$ using recursive iteration, which provides the desired answer but also yields an analog resource for $O(1)$ linear transformations of input signals/vectors using the resulting inverse matrix.

Moreover, the scalability of memristive crossbar arrays is a critical factor that influences their applicability in large-scale matrix computations. As the size of the crossbar array increases, issues related to interconnect resistance, crosstalk, and power distribution become more pronounced. Crosstalk and sneak paths can be reduced dramatically using the 1T1R approach, standing for 1 transistor 1 resistor. Although there are other techniques (1 selector 1 resistor 1S1R, and 1 diode 1 resistor 1D1R), the 1T1R is the standard approach to avoid crosstalk in the experimental setup. We will use this technique in our study.

In Sec. \ref{sec:sota} we provide a summary of previous works and a summary of the results obtained in this paper.  In Sec. \ref{sec:results} introduce all the paper's main results. In Sec. \ref{eq:algorithm} we introduce the algorithm we will be basing our results on as a dynamical system, and introduce the different variations to obtain different types of (pseudo-)inverses. In Sec. \ref{sec:spice} we discuss the simulation scheme and SPICE that we use to implement the inverses by simulating the crossbar. In Sec. \ref{sec:anres} we discuss analytical results and power consumption of the algorithm. In Sec. \ref{sec:applic} we discuss applications, and in particular we consider parameter learning of Gaussian distributions, and online learning for reservoir computing. Conclusions follow. All the derivations of the results of this manuscript are provided in the Appendices.

% Methods
%\section{Methods}

\section{Summary of previous works and analog methods}\label{sec:sota}

\subsection{Previous work}
Matrix operations are fundamental across many fields. For instance, in the traditional approach to matrix multiplication, multiplying two $N \times N$ matrices involves breaking one of the matrices into $N$ column vectors and performing $N$ matrix-vector multiplications. Each matrix-vector multiplication operation requires $O(N^2)$ operations, leading to an overall complexity of $O(N^3)$ for matrix multiplication. Significant research has focused on reducing this complexity. In 1969, Strassen was the first to break the $O(N^3)$ computational wall and introduced an algorithm that lowered the complexity from $O(N^3)$ to $O(N^{2.808})$ \cite{Strassen_1969}. Later, in 1978, Pan achieved a further reduction to $O(N^{2.796})$ \cite{Pan}. Coppersmith and Winograd were able to bring it down to $O(N^{2.496})$ \cite{Coppersmith_1982} (whose method settled to $O(N^{2.38})$ \cite{Coppersmith1987}), though reducing the complexity exponent below 2 has remained elusive. In 2003, Umans and Cohn \cite{Cohn2003} introduced a group theoretic method for matrix multiplication, followed by the result by Umans, Cohn,  Kleinberg, and Szegedy which we briefly discuss. They proposed that embedding matrix multiplication into the group algebra of a finite group could yield faster algorithms. This approach hinges on finding groups satisfying specific structural properties, termed the ``triple product property."
 Two conjectures arising from their work suggest that if proven, they could establish the matrix multiplication exponent as 2 (i.e. $O(N^2)$).  These algorithms still fundamentally rely on matrix-vector multiplications but optimize the total number of operations required through clever recursive strategies. Some comments are in order. First, as Higham puts it, ``to numerical analysts, matrix inversion is a sin" \cite{Higham2002}. For instance, it is not necessary to calculate the matrix inverse to solve $A x= \vec b$, and tailored algorithms are designed to solve this specific problem (although not changing the scaling in $N$ if not by a prefactor). This will be the case also for us later. Second, it is important to stress that the scaling in $N$ is not the only important quantity to keep into account. 

The conditioning number $ \kappa(A) $ of a matrix $ A $ quantifies how sensitive the solution (or inverse) is to perturbations in $ A $ and is defined as $\kappa(A)=\|A^{-1}\|$ where $\|\cdot \|$ is a generic matrix norm. A high conditioning number indicates that $ A^{-1} $ can amplify errors due to small perturbations in $ A $, potentially leading to less accurate solutions. Conversely, a low conditioning number indicates that $A^{-1} $ is less sensitive to such perturbations. In our case, the conditioning number affects the relaxation of the solution.
%When designing matrix inversion algorithms, particularly iterative or preconditioned methods, the conditioning number plays a critical role. Algorithms such as preconditioned iterative methods often rely on estimates of $ \kappa(A) $ to determine the optimal preconditioner. A well-chosen preconditioner can reduce the effective conditioning number of the problem, leading to faster convergence of the iterative solver. Error bounds and stability analysis for iterative methods often depend on the conditioning number. Algorithms like iterative refinement techniques \textcolor{red}{add....} explicitly consider $ \kappa(A) $ to refine the solution and improve accuracy. Effective preconditioners are designed to mitigate the effects of a high conditioning number, improving both the robustness and efficiency of iterative solvers. Techniques like incomplete factorizations or algebraic multigrid methods aim to reduce $ \kappa(A) $ effectively.
Solving linear algebra problems when dealing with matrices with high $ \kappa(A) $ requires careful algorithm selection and possibly preconditioning to ensure accurate results. Direct methods may be suitable for well-conditioned matrices or smaller problems where computational efficiency is less critical. For large or ill-conditioned matrices, iterative methods with appropriate preconditioning are often preferred. Algorithms should be chosen based on the specific properties of $ A $ (e.g., symmetric, positive definite) to ensure numerical stability and efficiency.
However, with the ever-increasing dimensions of matrices and the exponential growth in data, traditional methods and their improvements are becoming inadequate. Although quantum computers can achieve $O(\log(N) \kappa^2)$ in principle \cite{Harrow2009,Suba2019}, experimental tests of this algorithm have been obtained only for very small matrices. 

From the perspective of classical devices, the advent of parallel computing has shifted the focus towards developing efficient algorithms for large-scale, distributed matrix inversion. For instance, the von Neumann-Ulam algorithm is a probabilistic method for matrix inversion, leveraging the power of stochastic processes. The algorithm is based on iterative refinement and uses random sampling to approximate the inverse of a matrix \cite{forsythe1958, wu2021, sabelfeld2010}. The strength of the von Neumann-Ulam algorithm lies in its ability to handle large matrices more efficiently than deterministic methods via parallelization, especially when the matrix is sparse or has certain structural properties that can be exploited by the stochastic approach.
It is also worth noticing that certain ensembles of dense matrices have well-defined matrix inverses, something that has been noticed recently, and that can be calculated in $O(N^2)$ square time if nothing is known about row and column sums, and in $O(1)$ if the row and column sums are known \cite{Bartolucci_2021,Bartolucci2023,Bartoluccisparse}. It is also worth mentioning recent efforts on stochastic hardware accelerators for matrix inversion \cite{normal}.

To conclude, the Conjugate Gradient (CG) method \cite{Higham2002} is widely employed for solving symmetric positive definite linear systems, and is the closest iterative approach to the one we implement. Adaptations for positive semi-definite matrices $ A $ that are also $ s $-sparse extend its applicability. Given $ A \vec{x} = \vec{b} $, where $ A $ is $ s $-sparse and positive semi-definite, the CG method aims to find $ \vec{x} \in \mathbb{R}^N $ by minimizing the quadratic form $ |A\vec{x} - \vec{b}|^2 $, starting from

\begin{equation}
\vec{x}^{(0)} = \vec{0} \quad \text{and} \quad \vec{r}^{(0)} = \vec{b}
\end{equation}

For $ k = 0, 1, 2, \ldots $:
\begin{eqnarray}
\alpha_k &=& \frac{\vec{r}^{(k)T} \vec{r}^{(k)}}{\vec{p}^{(k)T} A \vec{p}^{(k)}}\\
\vec{x}^{(k+1)} &=& \vec{x}^{(k)} + \alpha_k \vec{p}^{(k)}
\vec{r}^{(k+1)} = \vec{r}^{(k)} - \alpha_k A \vec{p}^{(k)}\\
\beta_{k+1} &=& \frac{\vec{r}^{(k+1)T} \vec{r}^{(k+1)}}{\vec{r}^{(k)T} \vec{r}^{(k)}}\\
\vec{p}^{(k+1)} &=& \vec{r}^{(k+1)} + \beta_{k+1} \vec{p}^{(k)}
\end{eqnarray}
The algorithm terminates when a convergence criterion is met (e.g., residual tolerance or maximum iterations).
For a $ s $-sparse matrix $ A $, the Conjugate Gradient method operates with $ O(N s \kappa) $ time complexity, where $ N $ is the matrix size and $ \kappa $ is the condition number of $ A $ \cite{Saad2003}. This makes it suitable for large-scale problems where $ A $ is known to have a finite conditioning number. The drawback is that this method is not efficient on general-purpose processors, as it requires a sparse matrix-vector multiplication (SMVM) kernel \cite{Wiggers2007}, but can be implemented on a GPU \cite{Bolz2005}. In this paper, we ask whether a similar scaling can be obtained with an analog implementation on a cross-bar array.

\begin{figure}
    \centering
\includegraphics[width=\linewidth]{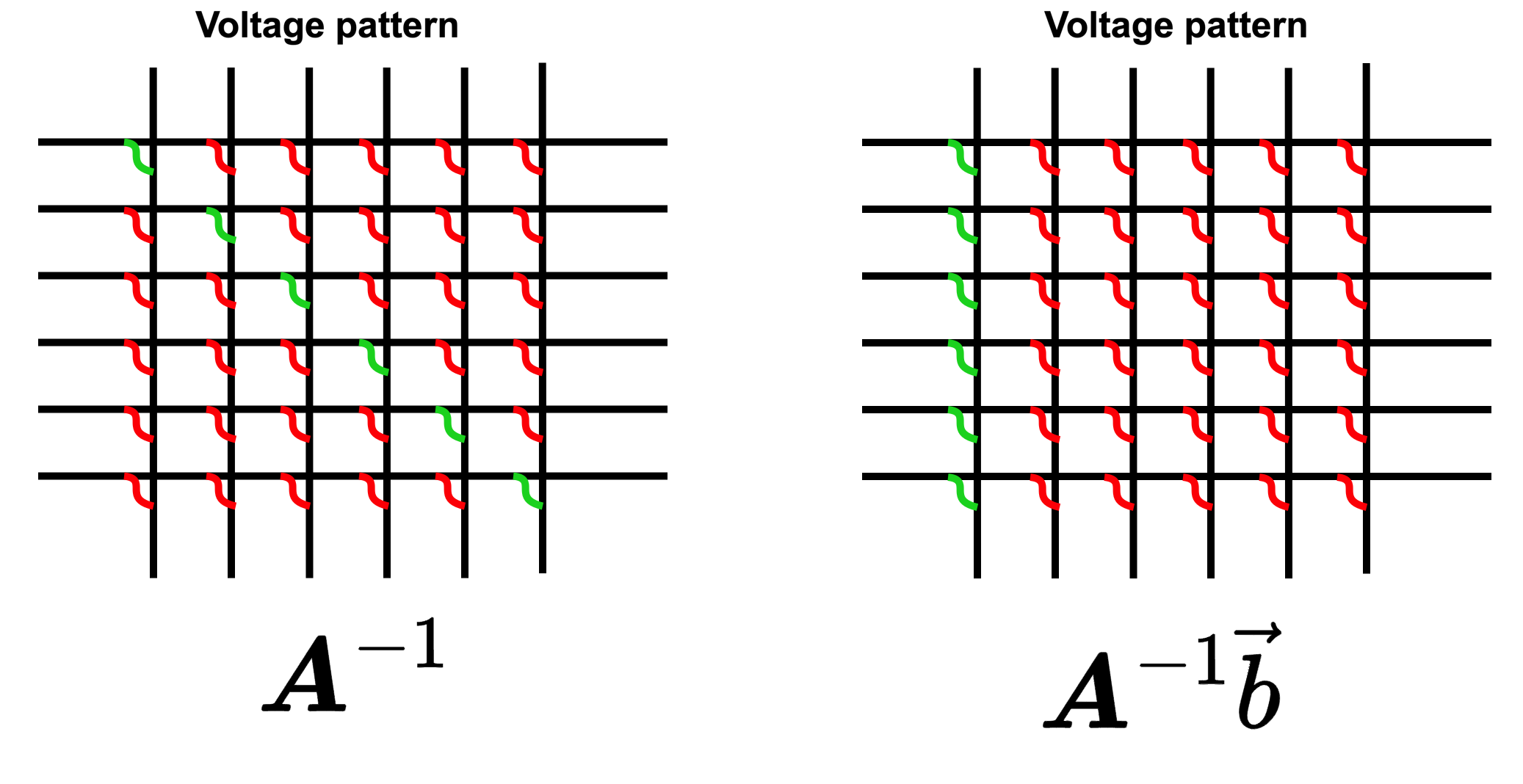}
    \caption{The figure above shows the constant voltage pattern used to solve the two linear problems on the crossbars. On the left, the constant voltage pattern is proportional to the diagonal matrix, and thus the bias is located 
 similarly on the crossbar elements. For the solution of the linear problem, we pick a column and place the vector $\vec b$. This implies that we can solve $N$ parallel instances of the same problem using this scheme.}
    \label{fig:pattern}
\end{figure}

\subsection{Summary of results}

This paper presents a novel approach to solving matrix equations using cross-point memristive arrays, commonly known as memristive crossbars. The main contributions and results of our study are summarized as follows:

\begin{itemize}
    \item We develop a method to leverage the properties of memristive devices to perform matrix inversion efficiently on crossbar arrays. The algorithm is described in Sec. \ref{eq:algorithm}. With this method, we can both solve for the matrix inverse when it exists, and we provided results for changing the approach to obtain the Moore-Penrose and Drazin pseudo-inverses. We also provide an algorithm to solve $N$ parallel $A\vec x=\vec b$ as a slight modification of the same algorithm. This can be done by changing the voltage patterns on the crossbar as shown in Fig. \ref{fig:pattern}. In particular, we have provided an online and offline implementation.   We observe that it is unfair to claim a 1-step matrix inversion, as several operations need to be implemented on an analog device to reach the solution. We introduce the number of operations (these are a combination of read and write operations) as a metric, and show that these scale linearly with the matrix size. We show that this method can be easily implementable on non-volatile devices, both for positive and negative matrices, and for volatile devices a backtracking method can be implemented to apply the inversion algorithm. An interesting observation is that there is a tradeoff between precision and power consumption using this algorithm. 

    \item  A key finding of our work is the proof of convergence for the proposed algorithm under noise and with the constraints on applied quantized (in time and values) of the applied voltages. We demonstrate that the algorithm reliably converges to the correct matrix inverse, ensuring its practical applicability and robustness. A summary is provided in Sec. \ref{sec:stability} and the proofs in the Appendices.

    \item  We validated our theoretical results through extensive simulations and on certain applications of interest. The empirical data supports our claims, showing that the memristive array-based method not only performs accurately but also offers considerable advantages in terms of speed and energy efficiency. We tested our algorithm on variational parameter learning, both online and offline, and tested a method for online reservoir computing in Sec. \ref{sec:applic}.

\end{itemize}
Despite these advancements, it is important to stress that algorithms whose inversion scales as $O(N^\alpha)$ with $\alpha\leq 2$ will have a scaling $O(N^2)$. The reason is that $O(N^2)$ is the price that one needs to pay for storing $A$ in the memory in the first place. This will be the case for us too\footnote{This fact is often overlooked in quantum computing analysis.}.

We provide the key results in the next section.

\section{Results} \label{sec:results}

\subsection{The main algorithm}\label{eq:algorithm}

The algorithm we employ for matrix inversion on a memristive crossbar array can be conceptualized as the emulation of a differential equation. This method uses voltage generators as drivers, applying voltages across the memristive elements to iteratively update their resistance values, effectively solving the system of linear equations. 

In a manner reminiscent of the Babylonian method for finding square roots, our approach iteratively refines an initial guess to converge on the desired solution. Specifically, the Babylonian method repeatedly adjusts the approximation of the square root by averaging it with the quotient of the number and the current approximation. Analogously, our algorithm incrementally updates the resistance matrix by applying a sequence of voltages, driving the system towards the matrix inverse.

We begin with a resistance matrix $R(t)$ and apply voltages $V_{ij}(t)$ across the memristors. We start with the simplest model of memristive dynamics \cite{Strukov_2008} to explain the basic idea, but this model can be implemented also by clamping the gating voltage of an array of diodes as in \cite{Stern_2021}. We will later discuss how this method is affected when window functions are introduced.

The dynamics of the system are governed by the differential equation:
\begin{equation}
\frac{dR_{ij}(t)}{dt} = -\alpha R_{ij}(t) + C_{ij} + \frac{R_d}{\beta} V_{ij}(t)
\end{equation}
where $R_{ij}(t)$ represents the resistance values, $\alpha$ and $\beta$ are constants, and $C_{ij}$ is a correction factor. 
Above, $V_{ij}(t)$ are voltages applied to the devices. Each memristive device operates in isolation during a given time step, allowing us to sequentially scan through the entire matrix. If a single device $(i,j)$ is operated at any time, then we can imagine that sequentially one scans through a matrix of devices. This type of operation can be implemented on a crossbar architecture with gating 1T1R as described before, to avoid the presence of sneak paths. This will be the assumption going forward.

We now describe the algorithm. Take the matrix of junction resistances $\mathbf{R}(t)$, some matrix $A$, and potentially some vector $\mathbf{b}$, with $\mathbf{R}, A \in \mathbb{R}^{N\times N}$ and $\mathbf{b} \in \mathbb{R}^{N}$. With proper characterization, memristive elements may be driven to evolve according to the general form
\begin{equation}
    \frac{dR_{ij}(t)}{dt} = \alpha(-\sum_{k}M_{ik}R_{ik}(t) + E_{ik}),
    \label{eq:gen_algo}
\end{equation}
or, simply in matrix notation,
\begin{equation}
    \frac{d\mathbf{R}(t)}{dt} = \alpha(-\mathbf{M}\mathbf{R}(t) + \mathbf{E}),
    \label{eq:gen_algo2}
\end{equation}
with $\mathbf{M}, E \in \mathbb{R}^{N\times N}$ and scalar $\alpha$ which we now choose. The steady state of the evolution is $R^* = \lim_{t\rightarrow\infty} R(t) = M^{-1} E$. Proofs of these statements are provided in App. \ref{app:basic}, and are obtained by inserting these expressions into the analytical solution of the dynamical system, which can be obtained by vectorizing the matrix system of equations. We note that the applied voltage requires the matrix multiplication $\mathbf{M}\mathbf{R}$, but this operation can be done in one step on the crossbar.

Then for different choices of $M, E$, we may implement different operations. If $A$ is known to be invertible, we may simply choose $M, E$ in the following ways:
\begin{itemize}
    \item $\mathbf{M = A, E = I}.$ $R^* = A^D$, the Drazin inverse of $A$, with $A^D = A^{-1}$ for invertible $A$.
    \item $\mathbf{M = A, E = B}.$ For a single $\mathbf{b}$, $B = \begin{bmatrix}\mathbf{b} & \mathbf{0}\end{bmatrix}$ and $R^* = \begin{bmatrix}\mathbf{x} & \mathbf{0}\end{bmatrix}$, where $\mathbf{x}$ is the solution of $A\mathbf{x} = \mathbf{b}$ (etc. for different choices of input column). We note also that for $\mathbf{b}_{1}, \ldots, \mathbf{b}_N$, choosing $B = \begin{bmatrix}\mathbf{b}_1 & \ldots & \mathbf{b}_N\end{bmatrix}$ would yield $R^* = \begin{bmatrix}\mathbf{x}_1 & \ldots & \mathbf{x}_N\end{bmatrix}$, i.e. a ``parallel" solution of $N$ problems.
\end{itemize}
For general $A$ we may compute the Gram matrix $G = A^T A$ in $\Theta(N^2 + N)$ using crossbar dynamics, and then choose $M, E$ in the following ways:
\begin{itemize}
    \item $\mathbf{M = G, E = A^T}.$ $R^*$ converges to the Moore-Penrose pseudoinverse of $A$, $A^\dagger$, for injective $A$, i.e. $A^\dagger = (A^TA)^{-1}A^T$.
    \item $\mathbf{M = G, E = A^T b}.$ This is included for completeness. $R^*$ converges to the least-norm solution of the linear system $A\mathbf{x} = \mathbf{b}$. In practice, one would drive $R^* = A^\dagger$ as above and then compute $A^\dagger \mathbf{b}$ as needed.
\end{itemize}

\textbf{Online formulation}. The fixed points of (\ref{eq:gen_algo}) are reached when $\mathbf{M}\mathbf{R} = \mathbf{R}\mathbf{M} = \mathbf{I}$. In particular, there exists a matrix $\mathbf{R}$ such that $\mathbf{M}\mathbf{R} = \mathbf{I}$ when $\mathbf{M}$ is square. In these cases we can instead right-multiply, in contrast to left-multiply, the input matrix $\mathbf{M}$ with the crossbar state $\mathbf{R}(t)$:
\begin{equation}
    \frac{d\mathbf{R}(t)}{dt} = \alpha(-\mathbf{R}(t)\mathbf{M} + \mathbf{E})
    \label{eq:online_algo}
\end{equation}
When this is true, only one crossbar is required for inversion iteration---the current state $\mathbf{R}(t)$ is both stored and used to compute the forcing for the next iteration, an example of so-called ``in-memory compute". The iteration circuit may therefore be implemented as a feedback loop (Figure \ref{fig:overview}), which will be useful for the online algorithmic applications we discuss.
The crossbar array enables us to implement and left- and right-multiply by applying bias on the left and right ends of the horizontal buses, respectively.

For square $A$, one may choose $M, E$ in (\ref{eq:online_algo}) in the following ways:
\begin{itemize}
    \item $\mathbf{M = A, E = I}.$ If $A$ is invertible, again $R^* = A^D$, the Drazin inverse of $A$, with $A^D = A^{-1}$ for invertible $A$. This is true because right and left inverses are identical for invertible square matrices.
    \item $\mathbf{M = AA^T, E = A^T}.$ $R^* = A^\dagger$, the Moore-Penrose pseudoinverse of $A$. 
\end{itemize}

%For a switch device, for $V_{ij}=0$, $C_{ij}$ is proportional to the resistance the memristor relaxes to. With the type of decay assumed above, $C_{ij}=\alpha R^{off}_{ij}$, guaranteeing that $R_{ij}^{eq}=R^{off}_{ij}$. It is easy to see that this is consistent with a switch, for which $R(x)=\Roff\ (1-x)+\Ron\  x$, whose internal memory parameter $x \in [0, 1]$ evolves according to $\dot x=-\alpha x+\frac{1}{\beta} V$, if $R_{d}=\Roff\ -\Ron\ .$

\subsection{SPICE implementation}\label{sec:spice}

\begin{figure*}
    \centering
    \begin{subfigure}{\textwidth}
        \centering
        \includegraphics[width=0.85\textwidth]{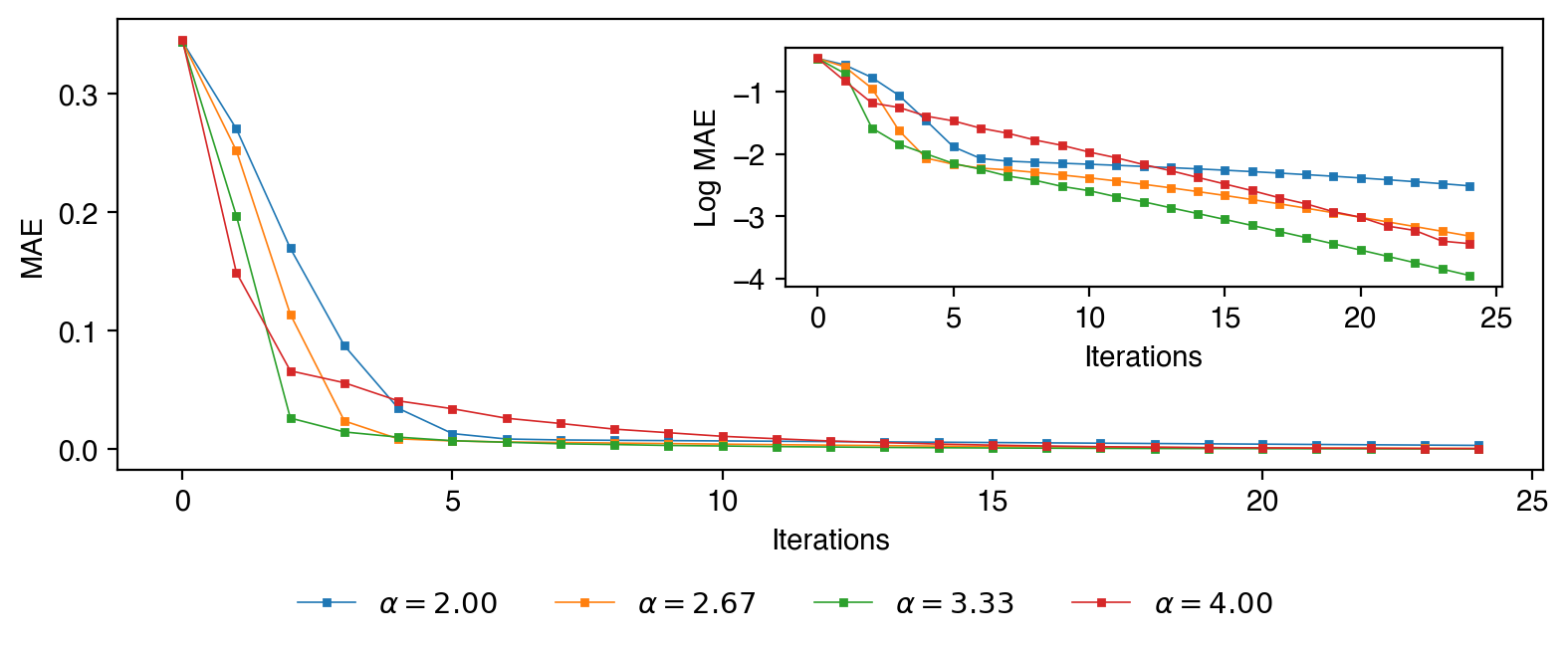}
    \end{subfigure}
    \begin{subfigure}{\textwidth}
        \centering
        \includegraphics[width=0.85\textwidth]{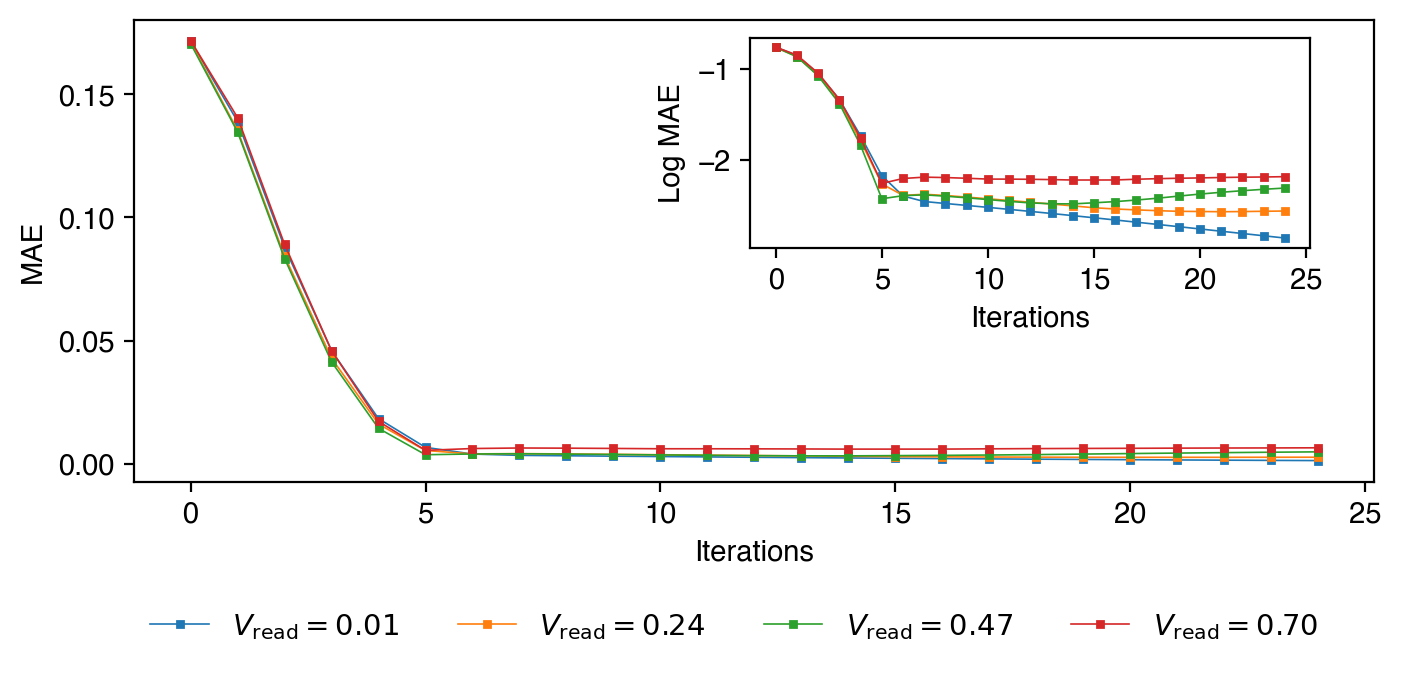}
    \end{subfigure}
    \caption{SPICE simulation of convergence performance of basic linear memristor implementation for various $\alpha$.  The mean absolute error between the crossbar state $R(t)$ and the inverse $A^{-1}$ converges exponentially to $\epsilon$ in $\alpha$. Joglekar-windowed HP memristors with $p=7$ are used in SPICE. \textbf{Inset:} Log convergence error (base 10).}
    \label{fig:basic_error}
\end{figure*}

% \begin{figure}
%     \centering
%     \includegraphics[width=0.5\textwidth]{spice/tau_space.png}
%     \caption{Convergence speed as a function of $\tau_\text{read}, \tau_\text{write}$.\textcolor{red}{fix axis labeling}}
%     \label{fig:tau_error}
% \end{figure}

\begin{figure}
    \centering
    \includegraphics[width=\linewidth]{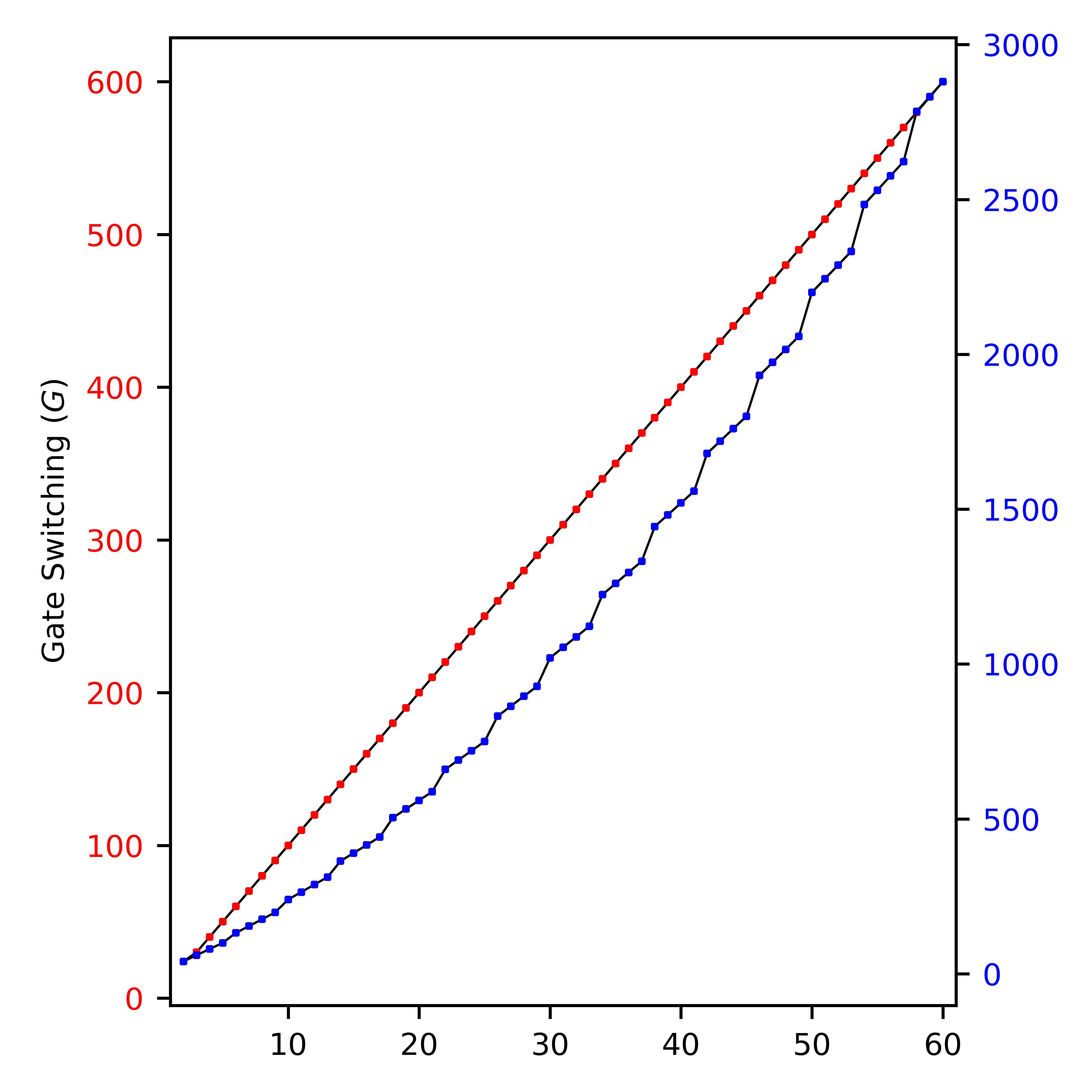}
    \caption{Total number of gate switching operations $G$ required for convergence to successfully invert matrices $\mathbf{A} \in \mathbb{R}^{N \times N}$. \textbf{Red.} Inverting random monotone matrices drawn from the distribution $A = \mathbf{I} - \frac{0.1}{N}U(N)$ results in a linear $G$ in $O(N)$. \textbf{Blue.} Inverting random matrices with fixed minimum eigenvalue 1.2 and $\kappa \approx 3$ reveals a weak dependence of iteration count on matrix size, giving a general operation complexity in $O(N^2)$.}
    \label{fig:step_complexity}
\end{figure}

\begin{figure*}
    \centering
    \begin{subfigure}{0.8\textwidth}
        \centering
        \includegraphics[width=\textwidth]{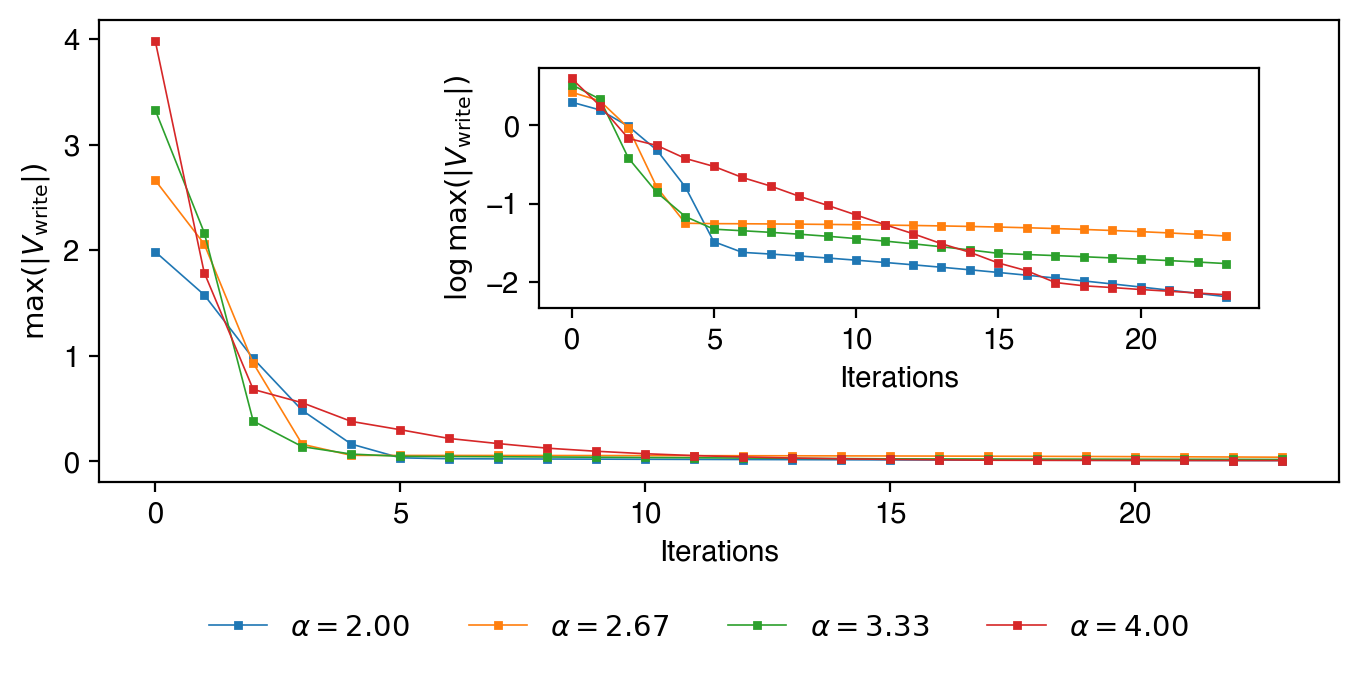}
        \caption{The maximum applied voltage $\max(V_\text{write})$ at each iteration. Here $N = 11$ and $\max(A) \approx 1.5$.}
    \end{subfigure}
    \begin{subfigure}{0.8\textwidth}
        \centering
        \includegraphics[width=\textwidth]{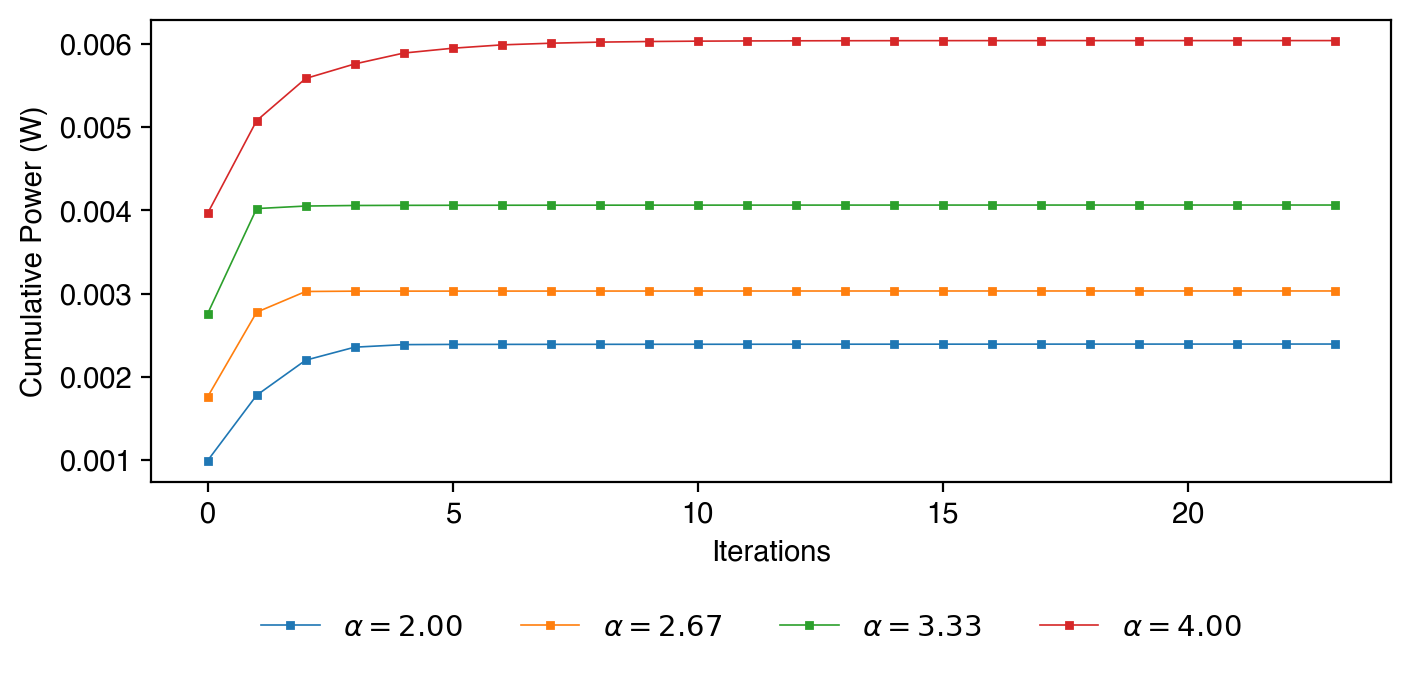}
        \caption{Cumulative power consumption as a function of iterations in watts. The consumption exponentially converges to an upper bound.}
    \end{subfigure}
    \caption{Power consumption curves for matrix $A$ with $N=11, \max(A) \approx 1.5$ for various $\alpha$.}
    \label{fig:vwrite_mag}
\end{figure*}

In this section, we discuss the methods and the obstacles that come from implementing such an algorithm on chip, and their possible solution. 

A standard crossbar requires memristive elements to be packaged in so-called 1T1R cells to allow for per-element read/write. In particular, the most efficient (na{\"i}ve) scheme restricts concurrent read/write access to diagonals of the crossbar matrix, meaning that $O(N)$ distinct operations of some duration $\tau$ are required to fully access the crossbar state.

In addition to this restriction, we wish to evaluate the effects of common practical issues like tuning error, read noise, and response time limitations in devices, which also introduce variability and delay. We investigate these issues in a SPICE implementation of our algorithm, performing a parameter sweep to capture the dynamics. Simulations are performed with the PySpice library, a wrapper over the Ngspice simulator.

We demonstrate general convergence in nonvolatile crossbar arrays of Joglekar-windowed TiO$_2$ memristor \cite{Strukov_2008,Joglekar_2009} model. Memristors are initialized with memory parameters uniformly drawn from the range $[0, 0.5]$. The voltage $V_{ij}(t)$ applied to every memristor takes the form $V_{ij}(t) = -\alpha (\mathbf{A}\mathbf{R})_{ij}(t - \tau) + \alpha \delta_{ij}$, where $\tau$ represents the delay introduced by component limitations. The input matrix $\mathbf{A} \in \mathbb{R}^{N \times N}$ has entries drawn from $I - \frac{0.1}{N} \cdot U(N)$, where $U(N)$ denotes a matrix with entries drawn uniformly from $[0, 1]$. We observe that the array state $R(t)$ converges exponentially (Figure \ref{fig:basic_error}) towards the inverse $A^{-1}$ with a speed proportional to $\alpha$, measuring the distance between $\mathbf{R}(t)$ and $\mathbf{A}^{-1}$ by computing the mean absolute error $\textsc{mae}(\mathbf{R}, \mathbf{A}^{-1}) = \frac{1}{N^2}\sum_{ij} |R_{ij}(t) - A^{-1}_{ij}(t)|$.

We assume that signals to the crossbar array are sent as square wave pulses of some duration $\tau$, with two relevant durations $\tau_\text{read}$ and $\tau_\text{write}$ for the respective stage within an iteration.

We next show the impact of read voltage $V_\text{read}$ on the convergence of the algorithm, as higher $V_\text{read}$ increases robustness to read noise but introduces an observer effect (Figure \ref{fig:basic_error}).

To gauge the time complexity of our algorithm, we fix a convergence error bound $\epsilon = \text{5e-3}$ and consider the number of distinct read/writes required to drive the crossbar $\textsc{MAE}$ to $\epsilon$ for increasing $N \in [2 \ldots 10]$. We note that in crossbars each read/write is (in our scaling assumption) $N$ atomic operations. Using this fact to obtain a loose analog to standard big-$O$ computational complexity $T$ for a given $N$. We also fix $\alpha = 15, \tau_\text{read} = \qty{4}{ms}, \tau_\text{write} = \qty{10}{ms}$. The resulting scaling is shown in Figure \ref{fig:step_complexity}.

We observe in simulation that for a certain class of diagonally-dominant asymmetric monotone matrices with fixed minimum eigenvalue 0.5, the iteration count converges to a constant $C$ at large $N$; the complexity of the algorithm is thus in $\Theta(3CN)$ or $O(N)$ for these matrices (Figure \ref{fig:step_complexity}). Thus the algorithm is in $\Omega(N)$. For more general matrices, we observe an average-case complexity in $O(N^2)$ for random matrices (Fig. \ref{fig:step_complexity}).

\subsection{Analytical results}\label{sec:anres}

\textbf{Computational complexity and advantages.} Iterative matrix inversion methods become indispensable in the solution of large linear systems and often prove to be the only option compared to direct methods. Such algorithms rely extensively on recursive applications of matrix multiplication, which is normally in $O(N^3)$. By contrast, we note that the dynamics of memristor crossbars allow for matrix multiplication in $O(N)$ via $N$ $O(1)$ matrix-vector products. We therefore expect some computational advantage in memristive iterative matrix inversion.

We precisely define an atomic crossbar operation in Appendix \ref{app:atomop}. By this definition, a single iteration of the algorithm has a complexity in $\Theta(3N)$. Our algorithm thus has an advantage as long as it converges to some error $\epsilon$ in less than $O(N^2)$ iterations. As with all iterative methods, this condition is predicated on $\kappa(\mathbf{A})$ and the extrema of the eigenspectrum of $A$, rather than its size $N$.

Consider an invertible matrix $\mathbf{A} \in \mathbb{R}^{N \times N}$ s.t. $A = Q\Lambda Q^{-1}$, with eigenvalues $\lambda_i = \{\lambda_1, \ldots, \lambda_N\}$, $\lambda_i \in \mathbb{C}$ and $\Lambda = \text{diag}(\lambda_1, \ldots, \lambda_N)$. Then the rate of convergence $\tau\mu$ is bounded only by $\min_i |\mathfrak{Re}(\lambda_i)|$. To see this, note from (\ref{eq:drazin_eigs}) that convergence depends on the evolution of the matrix exponential integral $I(t) = \int_0^t \alpha e^{-\alpha t \Lambda} \, dt$ as $t \rightarrow \infty$. Since $\Lambda$ is diagonal, each nonzero element $\lambda_i$ evolves independently according to $\int_0^t \alpha e^{-\alpha t \lambda_i} \, dt$. If we fix a convergence bound $\epsilon$, for a given $\lambda_i$ the integral reaches $\frac{1}{\lambda_i} \pm \epsilon$ at time $t_i = -\ln|\epsilon| / \alpha \lambda_i$. Thus over all independent convergence times $\{t_1, \ldots, t_N\}$, we see that $t_\text{max} = \text{argmax}_{i}\, t_i = \text{argmin}_{i}\, |\mathfrak{Re}(\lambda_i)|$, and $I(t_\text{max}) = \Lambda^{-1}$. Such a result is expected as matrices with near-zero eigenvalues are near-singular.

We then expect that for matrices with fixed minimum eigenvalues, the iteration count is independent of $N$, and thus that $\tau\mu \ll N$ for increasing problem sizes. Asymptotically, then, we expect that for matrices with amenable spectra, the time complexity is $\Theta(\tau \mu N) \in O(N)$ for the iterative portion of the algorithm. Indeed, we observe this effect in simulation (Figure \ref{fig:step_complexity}). However, an underlying assumption is that the evolution of the dynamics can be integrated stably by our system, determined by the condition number $\kappa(A)$. Thus the convergence rate depends also on the maximum eigenvalue of $A$, in the sense that above some $\kappa_\text{crit}(A)$, the error will never decrease below $\epsilon$.

\textbf{System size dependence.} In a physical crossbar with gated elements, one must also consider the finitely high impedance of nominally ``open" switches, which introduce a small amount of crosstalk between elements which depends on size. This effect introduces a dependence on $N$ that varies in the magnitude of the gating impedances, which we may call $R_\text{open}, R_\text{closed}$. In  simulation, we find that $R_\text{open} \approx 10^9\text{ }\Omega$ is necessary to invert large matrices with the algorithm (i.e. matrices with $N \gtrapprox 30$). We assume $R_\text{closed} \approx 0.05\text{ }\Omega$ without loss of generality, as different magnitudes may be absorbed into the rate coefficient $\alpha$.

\textbf{Optimal initialization.} The closed-form solution of the dynamics (\ref{eq:drazin_tr}) contains a transient term $e^{-\alpha t \mathbf{A}}\mathbf{R}(0)$, where $\mathbf{R}(0)$ is the initial matrix of resistances in the crossbar. This transient term must go to zero during convergence, introducing additional time complexity. However, we can simply apply a strong reset voltage to all elements in $O(1)$ such that $\mathbf{R}(0) \approx \mathbf{0}$ before the start of the iteration. Initializing $\mathbf{R}(0)$ in this manner results in exponential convergence, with a rate independent of $N$ when considering matrices with fixed spectral properties. This effect is expected given that the lower bounds on speed result from the eigenspectrum of $A$ and our convergence bound $\epsilon$, not $N$.

\textbf{Voltage requirements.} The element-wise magnitude of $V_\text{write}$ is proportional to the error, meaning that it reaches its maximum value early in iteration and decreases exponentially thereafter. We thus have an upper bound on $V_\text{write}$ for all steps a priori.

If we consider the maximum element $\max(\mathbf{M}) = \max_{ij} |M_{ij}|$ for some matrix $\mathbf{M}$, then we can find $\max(V_\text{write}(i))$ for a given iteration $i$. With $\mathbf{R}(0) \approx \mathbf{0}$ as described, clearly $V_\text{write}(0) \approx \alpha \mathbf{I}$ and therefore $\max(V_\text{write}(0)) \approx \alpha$, $\mathbf{R}(1) = \alpha \mathbf{I}$. Then $V_\text{write}(1) \approx -\alpha^2 \mathbf{A} + \alpha \mathbf{I}$ and $\max(V_\text{write}(1)) = |\alpha - \alpha^2\max(\mathbf{A})|$. This is the maximum required voltage for inversion; afterwards $\max(V_\text{write}(t)) \sim e^{-\alpha t / \kappa(\mathbf{A})}$. The maximum voltage required to invert $A$ is therefore proportional to the maximum element of $\mathbf{A}$ and the speed with which we wish to invert it.

\textbf{Power consumption.} In our simulations, the power consumption is well-approximated by considering the iteration sequence $\mathbf{R}_0, \mathbf{R}_1, \mathbf{R}_2, \ldots$ and the input matrix $\mathbf{A}$. We may then compute $\mathbf{V}_i = -\alpha(-\mathbf{A}\mathbf{R}_{i-1} + \mathbf{I})$, taking $\mathbf{V}_0 = \alpha\mathbf{I}$ given $\mathbf{R}_0 \approx \mathbf{0}$. The cumulative power consumption at some iteration $i$ is then $\sum_{i} \sum_{jk} \mathbf{V}_{jk}^{2} / \mathbf{R}_{jk}$.

Because the algorithm converges exponentially to the state $\mathbf{A}^{-1}$, the elementwise magnitude of $\mathbf{V}$, which may be thought of as an error or correction term, decreases exponentially. Thus the cumulative power consumption converges exponentially to an upper bound, as shown in Fig. \ref{fig:vwrite_mag}.

\subsection{Stability and Error}\label{sec:stability}

%\textcolor{red}{Stability: time-delay \ref{app:delayana}, stochasticity \ref{app:perturbationdrazin}, mention results from appendix here}

In this section, we present a detailed analysis of the algorithms used in our study, highlighting the results presented in the appendix for further insights.
One comment to make is that purely analog (continuous time) control does not exist. In practice, voltage is controlled in steps, and thus in many ways, the effective algorithm being implemented is a delayed differential equation. 
We consider the reduced dynamics of the crossbar inversion algorithm, a gradient flow that follows the dynamics
\begin{equation}
    \frac{d\mathbf{R}(t)}{dt} = -\alpha \mathbf{A} \hat{\mathbf{R}}(t-\tau) + \alpha \mathbf{I}\label{eq:dode}
\end{equation}
where $\hat{\mathbf{R}}(t)$ represents the estimated value of the crossbar state from a read and $\tau \geq 0$ is a delay in the estimation of the state $R$. In systems where the error in $\hat{\mathbf{R}}(t)$ grows with time due to volatility ($\tau > 0$), such delay-induced error results in oscillations in the evolution of the crossbar state about the true inverse. 

We can also consider the effects of observational noise ($\xi_O$) and process noise ($\mathbf{\xi}_P$), writing
\begin{align}
    \frac{d\mathbf{R}(t)}{dt} = -\alpha \mathbf{A}(\mathbf{R}(t-\tau) + \xi_O) + \alpha \mathbf{I} + \mathbf{\xi}_P 
\end{align}

While process noise $\mathbf{\xi}_P$ can be averaged out or predicted with standard techniques, $\xi_O$ enters as a nonlinear term. Its influence is proportional to the spectral radius $\rho(A)$. In practice, however, if the RMS noise magnitude is negligible compared to the magnitude of $V_\text{read}$, $\mathbf{\xi}_P$ may be ignored.

More specifically, the appendix contains detailed derivations and proofs of the key results used in our analysis. 
 This proof demonstrates that under certain conditions, the algorithm converges to the true inverse of the matrix, leveraging the inherent properties of memristive devices, see App. \ref{app:basic}) for a detailed proof. 
The convergence proof is central to our algorithmic analysis, as it establishes the foundational guarantee that our method reliably produces the correct inverse under specified conditions. This proof utilizes techniques from dynamical systems theory to show that the feedback mechanism used in the memristive array ensures stability and convergence.

In App. \ref{sec:stochan}, the error analysis provides insights into how device imperfections affect the algorithm's performance asymptotically. By modeling the variability in memristive devices, we derive upper bounds on the error, which are crucial for understanding the practical limitations of our approach, including the  Moore-Penrose and Drazin inverses. These two converge to the same matrix if the matrix is invertible, but the response to noise is different. If we assume that the right-hand side is perturbed by an i.i.d. noise $\xi(t)$ such that $\sigma$ is the strength of the noise $\langle \xi^2\rangle=\sigma^2$, then we obtain the elementwise error is given by

\begin{eqnarray}
&&\langle \mathbf{R}^2(t)\rangle-\langle \mathbf{R}(t)\rangle^2=\nonumber \\
&&\hspace{2cm}\begin{cases}    \frac{2\sigma^2}{\alpha}\int_0^te^{-s  \mathbf{A}}\,ds\ & \textit{Drazin}\\
    \frac{2 \sigma^2}{\alpha}\int_0^t e^{- s  \mathbf{A^t A}}\,ds & \textit{Moore-Penrose}
     \end{cases}\nonumber 
\end{eqnarray}
which, as we can see, depends on time and is element dependent, but it is proportional in both cases to $2\sigma^2/\alpha$. This implies that noise can be mitigated by running the analog at a slower pace, determined by the constant $\alpha$.

In addition, we have studied the case in which volatility is present in the devices, and delays are present in the algorithm. Specifically,
a comprehensive stability analysis is presented in App. \ref{app:delayana}, quantifying the stability to non-idealities and volatility in the control, and on the accuracy of the computed inverse, both for Moore-Penrose and Drazin inverses using stochastic analysis. In particular, we assume that the applied voltage has a delay $\tau$ with respect to the readout. We show the classification of the instability according to the location of the pole. In particular, for small values of the delay and for volatile devices stable oscillations occur. The oscillations become unstable for longer delays.

%\subsection{``Online" formulation}

% \subsection{Numerical implementation}

% \textbf{Stability for artificial models.}

% \textbf{SPICE models.}

\section{Applications}\label{sec:applic}
%\subsection{Katz centrality}
%\input{secs/katz}
We now describe two applications where we can implement matrix inversion both offline and online. To be clear, in the offline version, all data is at the beginning of the algorithm, while in the online version, data arrives over time. In both cases we describe below, we can implement the matrix inversion.
\subsection{Variational parameter learning}
\begin{figure*}
    \centering
   \includegraphics[width=0.99\textwidth]{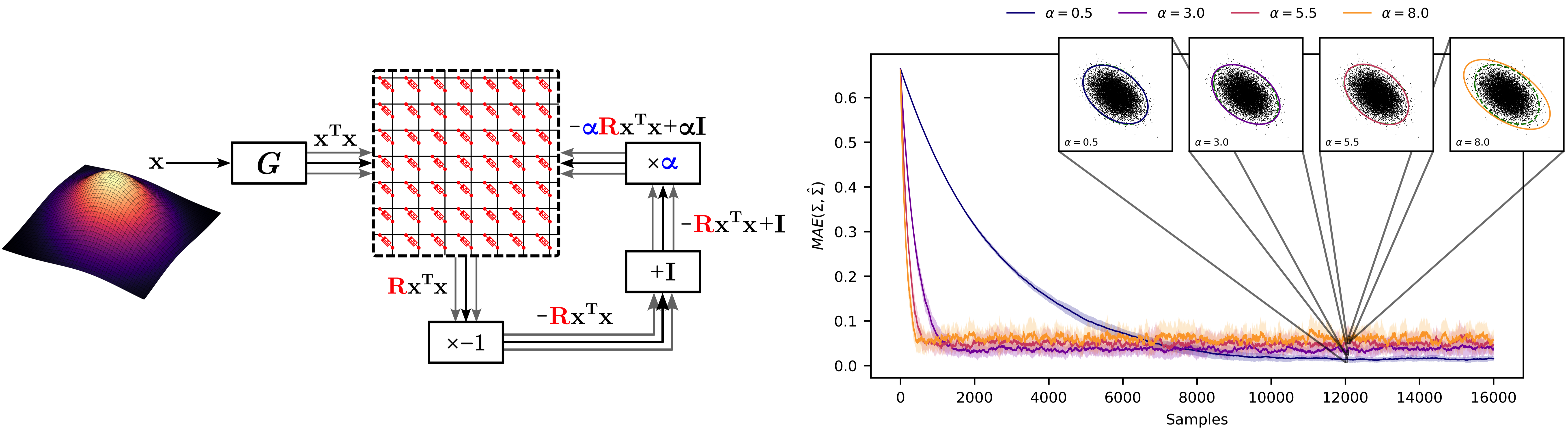}
    \caption{With $M=1$---a single sample per iteration---we already achieve a good estimate $\hat{\Sigma}$ for $\Sigma$ when the crossbar state evolves according to (\ref{eq:online_varinf}). The real-time equivalent of convergence is about 14 seconds, but it should be noted that this estimate evolves from a complete lack of knowledge of $\Sigma$. ``Online" usage after initial convergence should converge very quickly, with only incremental updates to the crossbar state. The \fb{right} plot gives MAE curves for online inference with select values of $\alpha \in [0.5, 8]$, showing that higher $\alpha$ yield faster convergence to an estimate, albeit with larger fluctuations and less accuracy. The true distribution is marked by a dashed green line.}
    \label{fig:varinf}
\end{figure*}
As an application of the analog inverse method, we consider the parameter learning of a Gaussian distribution from samples \cite{var0,var1}. The Gaussian distribution we consider can be parametrized in the form
\begin{eqnarray}
    p_\Sigma(\vec x)&=&\frac{1}{Z(\Sigma)} e^{-\frac{1}{2} \vec x^t \Sigma \vec x}
\end{eqnarray}
We assume that we want to learn the parameters $\Sigma_{ij}$ from samples of the distribution, $x_i^k$ where $k$ represents the sampled values and $i$ are the vector elements. 
A common method used to infer these parameters is the KL divergence minimization via gradient descent (see App. \ref{sec:paramlearnapp}), which leads to the equation
\begin{eqnarray}
    \frac{d(\Sigma_t)_{ab}}{dt}=\xi(\langle x_a x_b\rangle_{emp}-(\Sigma_t^{-1})_{ab})
\end{eqnarray}
or rather, its Euler discretization
\begin{eqnarray}
    \label{eq:varinfgrad}
    (\Sigma_{t+1})_{ab}=(\Sigma_{t})_{ab}+ dt\,\xi(\langle x_a x_b\rangle_{emp}-(\Sigma_t^{-1})_{ab}),
\end{eqnarray}
where $\Sigma_t$ is the matrix at time $t$, $\xi$ is the learning rate and $\langle x_a x_b\rangle_{emp}$ is the empirical average, e.g.
\begin{eqnarray}
    \langle x_a x_b\rangle_{emp}=\frac{1}{M} \sum_{k} x_a^k x_b^k,
\end{eqnarray}
where $M$ is the total number of samples. With some sample matrix $X \in \mathbb{R}^{M \times D}$ at each iteration, this is also the averaged Gram matrix $\langle x_a x_b\rangle_{emp} = \frac{1}{M} X^T X$.

Notice that the gradient flow in (\ref{eq:varinfgrad}) is essentially computing the quantity $\hat{\Sigma}_{ij} = \left\langle \left(\frac{1}{M} X^T X\right)^{-1}\right\rangle_{emp}$. Because the estimator $\hat{\Sigma}$ is square and invertible, we may thus slightly modify (\ref{eq:online_algo}) to the form
\begin{equation}
    \dot{R}(t) = \alpha(-R(t)\langle x_a x_b\rangle_{emp} + I)
    \label{eq:online_varinf}
\end{equation}
i.e. the ``online" form of (\ref{eq:gen_algo}), such that $\langle R^* \rangle = \Sigma_{ij}$ with large enough $M$ and small enough $\alpha$. We find that in practice, excellent performance can already be achieved with $M = 2, \alpha=1$ (Figure \ref{fig:varinf}), making this approach a viable candidate for an online implementation of variational inference. A crossbar driven to such an estimator of $A$ can then be used to transform another white noise source into a similar distribution, a useful operation in applications like noise cancellation.

\subsection{From Offline to Online Reservoir Computing}
\begin{figure*}
    \centering
    \includegraphics[width=0.4\textwidth]{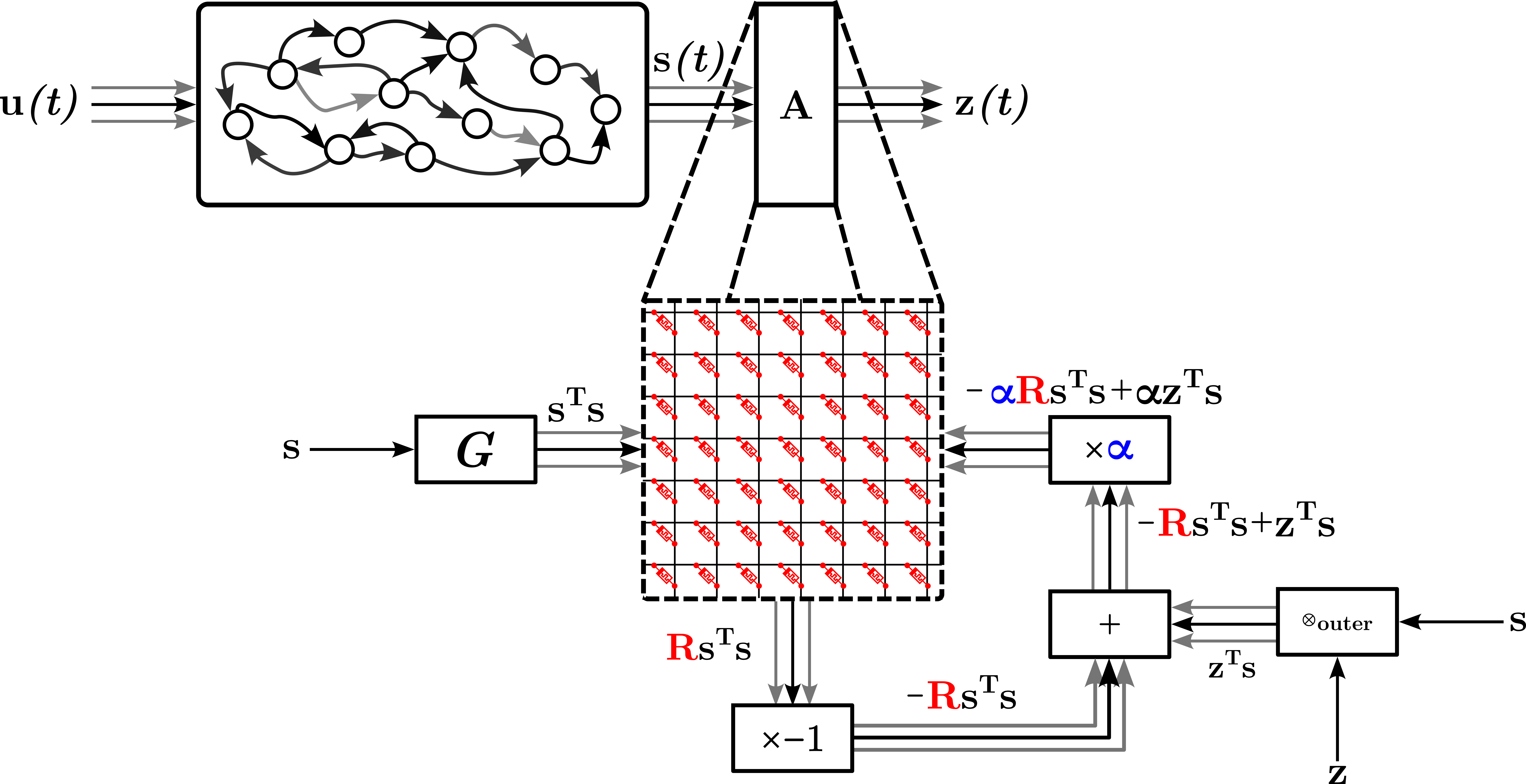}
    \raisebox{-0.2\height}{\includegraphics[width=0.58\textwidth]{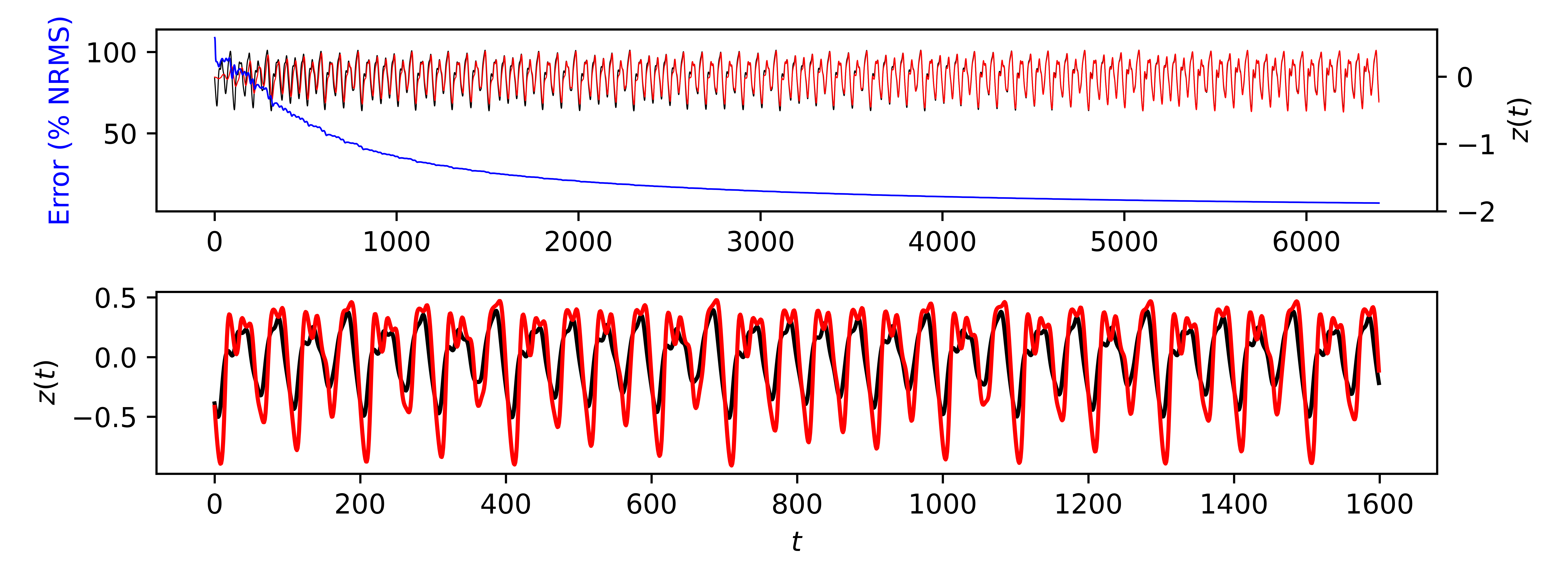}}
    \caption{Online reservoir computing on a constant stream of sampled $(\mathbf{s}, \mathbf{z})$ from an ESN. Training error converges to 1.679\% as samples arrive. In the left figure, we plot the evolution of the error as a function of time, using an online learning scheme for a Mackey-Glass time series. At the beginning, the system is not tuned to fit the time series. The regression matrix is being learned as data arrives, and the cumulative NRMSE goes down accordingly. We can see a zoom of the final fit in the right plot, for the prediction mapping $\mathbf{s}(t) \mapsto \mathbf{z}(t + 5)$.}
    \label{fig:online_lstsq}
\end{figure*}

The previous section described an online, iterative computation of the matrix inverse. We now consider its potential application to on-chip reservoir computing \cite{res0,res1}. In particular, we ask if this algorithm can be used for online learning---and computation---of the transformation on the reservoir via iterative pseudoinverse computation.

We give a brief overview of reservoir computing. A reservoir computer is a driven dynamical system that is trained to transform an input driving signal into an output trajectory. Given an input driving signal $\mathbf{u}(t)$ and target trajectory $\mathbf{z}(t) \in \mathbb{R}^O$, and assuming a system with state readout $\mathbf{s}(t) \in \mathbb{R}^D$ and sufficiently nonlinear $F$ that evolves according to
\begin{equation}
    \dot{\mathbf{s}}(t) = F(\mathbf{s}(t)) + \mathbf{u}(t)
\end{equation}
we ask for a linear transformation $\mathbf{A} \in \mathbb{R}^{O \times D}$ such that $\mathbf{As}(t) \approx \mathbf{z}(t)$.

Commonly, the output states and corresponding target trajectories are discretely sampled at $T$ times such that we have matrices $\mathbf{S} \in \mathbb{R}^{T \times D}, \mathbf{Z} \in \mathbb{R}^{T \times O}$. Then by standard linear least-squares arguments, an approximation for the transformation $\mathbf{A}$ is given by the Moore-Penrose pseudoinverse, i.e.
\begin{equation}
    \argmin_\mathbf{A} ||\mathbf{A}\mathbf{s}(t) - \mathbf{z}(t)||^2 = [(\mathbf{S}^T\mathbf{S})^{-1}\mathbf{S}^T\mathbf{Z}]^T
    \label{eq:reservoir_est}
\end{equation}

We put the approximation in the form above $([\ldots]^T)$ to ask if an online computation of (\ref{eq:reservoir_est}) can be achieved with our method. By similar arguments as in the previous sections, we arrive at
\begin{equation}
    \dot{\mathbf{R}}(t) = \alpha(-\mathbf{R}(t)(\mathbf{s}^T\mathbf{s} + \lambda\mathbf{I}) + \mathbf{z}^T\mathbf{s})
    \label{eq:online_pinv}
\end{equation}
where $\mathbf{s}\in\mathbb{R}^{1 \times D},\mathbf{z}\in\mathbb{R}^{1 \times O}$ are now single samples of output states and target trajectories respectively, and $\lambda$ is the familiar ridge regression control parameter. Given a suitably distributed sequence of paired ($\mathbf{s}, \mathbf{z}$), this iterative least-squares regression indeed converges to (\ref{eq:reservoir_est}) in an online fashion (Figure \ref{fig:online_lstsq}). We use the paired-crossbar architecture discussed in Appendix \ref{sec:mixed_sign} to address the mixed signs in the estimator $\hat{A}$.

The algorithm also works when the desired mapping is rectangular, i.e. $A \in \mathbb{R}^{O \times D}$ and $O \ll D$. In such cases the gating transistors on the lower $D - O$ rows of the crossbar can be switched off for the full duration of execution, effectively zeroing those matrix elements. We use this fact to map the full reservoir state to a far smaller output dimensionality in an online fashion.

We note that (\ref{eq:online_pinv}) converges under similar assumptions as (\ref{eq:online_varinf}), i.e. that the samples $(\mathbf{s}, \mathbf{z}) \leftarrow \text{span}(A)$ arrive roughly i.i.d. in time. To see this, we fix $\lambda = 0$ and note that a fixed point of (\ref{eq:online_pinv}) exists when $\mathbf{R^*}(\mathbf{s}^Ts) = \mathbf{z}^T\mathbf{s}$. Suppose that the pairs $(\mathbf{s}, \mathbf{z})$ are drawn uniformly and randomly from the span of $\mathbf{A}$, such that we have random variables $\langle\mathbf{s}^T\mathbf{s}\rangle, \langle\mathbf{z}^T\mathbf{s}\rangle$. Assuming the existence of $\mathbf{A}$, the fixed point condition thus becomes $\mathbf{R^*}\langle\mathbf{s}^Ts\rangle = \mathbf{A}\langle\mathbf{s}^T\mathbf{s}\rangle$, and we see that a fixed point is reached when the time average of the outer product $\langle \mathbf{s}^T\mathbf{s}\rangle$ becomes stationary. This requirement also becomes clear if one notes that (\ref{eq:online_pinv}) essentially implements a first-order recursive least squares (RLS) filter.

These conditions can only be met by a reservoir which is sufficiently chaotic. In these systems, the Lyapunov exponent is large enough to effectively decorrelate consecutive samples of trajectories in time. We employ an echo state network (ESN) in ReservoirPy as the reservoir, with $N = 600$ neurons, leakage rate $\alpha = 0.99$, spectral radius $\rho = 0.99$, and reservoir connectivity $p = 0.7$. We train the reservoir on a Mackey-Glass sequence prediction task, preprocessing the input and target sequences by learning the mapping $\mathbf{s}(t) \mapsto \mathbf{u}(t + \tau), \tau \in [1 \ldots 5]$. For Mackey-Glass, the control parameters $r=0.2, \gamma=0.1, \tau=17, n=10, x_0=0.1$ are used.

% Discussion
\section{Discussion}

The advent of parallel computing has shifted the focus towards developing efficient algorithms for large-scale, distributed matrix inversion. In this paper, we present several key results that highlight the advantages of utilizing memristive crossbar arrays for matrix inversion.

Our main algorithm emulates a differential equation to iteratively update the resistance values in a memristive crossbar array, effectively solving the system of linear equations. We demonstrate that our algorithm converges reliably to the correct matrix inverse, leveraging the inherent properties of memristive devices. Moreover, we have provided error estimates in the presence of noise, and shown that the algorithm is stable in the present of delays, relevant to the case of volatile devices.

We validated our theoretical results through extensive simulations, showcasing the practical applicability and robustness of our approach. The empirical data supports our claims, showing that the memristive array-based method not only performs accurately but also offers considerable advantages in terms of speed and energy efficiency. Our simulations indicate that the cumulative power consumption converges exponentially to an upper bound, demonstrating the energy efficiency of our approach. The simulations were conducted using SPICE, ensuring realistic device behavior and providing a strong foundation for the practical implementation of the algorithm. 

One of the key advantages of our method is its scalability. The number of operations required for convergence scales linearly with the matrix size, which is a significant improvement over traditional methods that often exhibit quadratic or higher-order complexity. This makes our approach particularly suitable for large-scale problems where conventional methods become computationally prohibitive.

We also highlight that our algorithm can handle both positive and negative entries in the matrix, which is a crucial capability for real-world applications. Theoretical results with noise show that our algorithm maintains stability and converges to the true inverse of the matrix under certain conditions. We provide upper bounds on the error, demonstrating that noise can be mitigated by adjusting the algorithm's pace (e.g. the applied voltage).

Furthermore, we provide analytical results that underpin the convergence and stability of our algorithm, reinforcing the reliability of our approach. The ability to handle observational and process noise with analytical guarantees adds to the robustness of our method.

Our analysis extends to non-invertible matrices, where we successfully apply our method to compute the Moore-Penrose inverse and the Drazin inverse. This flexibility underscores the robustness of our approach and its applicability to a wide range of linear algebra problems.

Moreover, in terms of implementation schemes, we demonstrate that our algorithm can be adapted to solve both for $A^{-1}$ and for $N$ parallel instances of the type $A^{-1} \vec b$
 when the matrix is invertible, offering versatility in its application. This dual capability allows for broader use in various computational contexts.

The numerical simulations we provided corroborate our theoretical findings, showing that our method performs well even with realistic device imperfections. The results indicate that memristive crossbar arrays can be effectively used for matrix inversion, providing a scalable, energy-efficient alternative to traditional methods.

Most importantly, the present manuscript shows that the use of crossbar arrays with memristive devices can be used both for online and offline matrix inversion. We have tested these schemes on two applications, i.e. parameter learning of multi-variate Gaussian distributions and reservoir computing. Specifically, we have shown that our method rapidly converges to the optimal solution.

In summary, our study demonstrates that memristive crossbar arrays offer a promising approach for efficient matrix inversion, with potential applications in various fields requiring large-scale linear algebra computations. The results presented in this paper provide a strong foundation for further research and development in this area. In our future work, we will focus on the physical implementation of the algorithm, exploring its practical deployment in more realistic case studies, and on-chip.

% Conclusion
%\section{Conclusion}

% Acknowledgments
\section*{Acknowledgments}
The authors acknowledge the support of NNSA for
the U.S. DoE at LANL under Contract No. DE-AC52-
06NA25396, and Laboratory Directed Research and Development (LDRD). FC and JL were financed via DOE LDRD grant 20240245ER, while FB via Director’s Fellowship. FB and JL also gratefully acknowledge support from the Center for Nonlinear Studies at LANL. The authors are indebted to A. Jayakumar for some enlightening discussions on parameter learning.

\bibliography{references}

\clearpage
\appendix
\onecolumngrid
\section{Basic derivations}\label{app:basic}

Each memristor follows an equation of the form
\begin{equation}
    \frac{dR_{ij}}{dt}=-\alpha R_{ij}+C_{ij}+V_{ij}(t)\label{eqapp:eq1}
\end{equation}
and we ask for solutions such that $R_{ij}^*=(\mathbf{A}^{-1})_{ij}$. Let us assume that $0\leq R_{ij}\leq \infty$. The values of $C_{ij}$ are the natural resistance values at equilibrium.

Let us choose $\mathbf{V}=\alpha( -\mathbf{A}+\mathbf{B}) \mathbf{R}-\mathbf{C}+ \mathbf{M}$, and let $\mathbf{A}$ be a monotone  matrix, e.g. $\mathbf{A}^{-1}$ be positive element-wise. Choosing $\mathbf{M}=\alpha^*\mathbf{I}$ provides the inverse of $\mathbf{A}$, while choosing $\mathbf{M}=\text{diag}(\vec b)$ provides the solution of $\mathbf{A}\vec x=\vec b$ if $\mathbf{A}$ is invertible.

Let us consider
\begin{equation}
    0=-\alpha \mathbf{R}+\mathbf{C}+ \alpha (-\mathbf{A}+\mathbf{B}) \mathbf{R}-\mathbf{C}+\mathbf{M}
\end{equation}
It follows that 
\begin{equation}
    \alpha (\mathbf{I}+\mathbf{A}-\mathbf{B})\mathbf{R}=\mathbf{M}
\end{equation}
and thus
\begin{equation}
    \mathbf{R}=-\frac{1}{\alpha}(\mathbf{I}+\mathbf{A}-\mathbf{B})^{-1} \mathbf{M}.
\end{equation}
If we choose $\mathbf{B}=\mathbf{I}$, then
\begin{equation}
    \mathbf{R}=\frac{1}{\alpha}\mathbf{A}^{-1} \mathbf{M}.
\end{equation}
For $M=\alpha I$, then the solution is directly the matrix inverse. Let us choose $M_{ij}=\alpha*b_i \delta_{j1}$. Then
\begin{equation}
    R_{rt}=\sum_{k} (A^{-1})_{rk} b_k \delta_{t1}
\end{equation}
Thus, the memristive devices $R_{i 1}=x_i$.

Let us consider the vectorized version of the differential equation. To avoid doubts, want to show that we can write the matrix solution as we would write the vector solution of the differential equation.  We define $ \text{vec}(M)$ to be the vector with the columns of $M$ stacked. Let us call $\vec r=\text{vec}(R)$ and $\vec e=\text{vec}(I_N)$. Then, using the fact for the matrix multiplication we have
\begin{equation}
    \text{vec}(AR) =(I_N\otimes A) \vec r\equiv \mathcal A \vec r.
\end{equation}
and the differential equation can be written in the form
\begin{equation}
    \frac{d}{dt} \vec r=-\alpha \mathcal A \vec r+\alpha \vec e.
\end{equation}

Let us briefly discuss on the stability.  We consider the following system of gradient descent dynamical equations
\begin{equation}
    \frac{dx_{ij}}{dt}=-\frac{\partial}{\partial x_{ij}} \mathcal L(A,x)
\end{equation}
where 
\begin{eqnarray}
    \mathcal L(A,x)&=&\frac{1}{2}\sum_{ijlm}(\sum_{k} A_{ik} x_{kj}-\delta_{ij})A^{-1}_{jl}(\sum_{r} A_{lr} x_{rm}-\delta_{lm}).\nonumber 
\end{eqnarray}
from which we obtain, for $\mathbf{A}$ symmetric, invertible and positive, that
\begin{eqnarray}
    \frac{\delta\mathcal L}{\delta x_{ij}}=- \sum_k A_{ik} x_{kj}+\delta_{ij}.
\end{eqnarray}
As a result, the continuous-time algorithm is stable.

\subsection{Drazin inverse}
In the case when we choose $\mathbf{M} = \alpha \mathbf{I}$, we also see that the matrix $\mathbf{R}$ converges to $\mathbf{A}^{-1}$ by assuming a diagonalizable $\mathbf{A}$ and solving for the closed form solution of the matrix differential equation $\dot{\mathbf{R}} = -\alpha \mathbf{A}\mathbf{R} + \alpha \mathbf{I}$, where we discard the $\mathbf{B}$ and $\mathbf{C}$ terms by similar steps as above and substitute $\mathbf{M}$. The solution is then
\begin{align}
    \mathbf{R}(t) &= e^{-\alpha t\mathbf{A}}\mathbf{R}(0) + \int_0^t e^{(t-\tau)(-\alpha \mathbf{A})} \alpha \mathbf{I}\,d\tau \\
    &= e^{-\alpha t\mathbf{A}}\mathbf{R}(0) + \alpha\int_0^t e^{-\alpha(t-\tau)\mathbf{A}}\,d\tau \label{eq:drazin_tr}
\end{align}
For positive semi-definite $\mathbf{A}$ (note $-\alpha$) the constant term vanishes and the limiting value $\mathbf{R}_{\infty}$ becomes
\begin{equation}
    \lim_{t \rightarrow \infty} \mathbf{R}(t) = R_\infty = \alpha\int_0^\infty e^{-\alpha t\mathbf{A}}\,dt
\end{equation}

Now we assume that $\mathbf{A}$ is symmetric. We can then diagonalize it via $\mathbf{A} = \mathbf{P}\mathbf{D}\mathbf{P}^{-1}$, yielding
\begin{align}
    \mathbf{R}_\infty &= \alpha\int_0^\infty \mathbf{P}e^{-\alpha t\mathbf{D}}\mathbf{P}^{-1}\,dt \\
    &= \alpha \mathbf{P}\left(\int_0^\infty e^{-\alpha t\mathbf{D}}\, dt\right)\mathbf{P}^{-1} \label{eq:drazin_eigs} \\
    &= \alpha \mathbf{P}\left(-\frac{1}{\alpha}\mathbf{D}^{-1}e^{-\alpha t\mathbf{D}}\Bigg|^\infty_0\right)\mathbf{P}^{-1} \\
    &= \alpha \mathbf{P}\left(-\frac{1}{\alpha}\mathbf{D}^{-1}(-I)\right)\mathbf{P}^{-1} \\
    &= \mathbf{P}\mathbf{D}^{-1}\mathbf{P}^{-1} = \mathbf{A}^{-1}
\end{align}
where the improper integral converges because $\mathbf{A}$ is positive semi-definite. We see then that this flow algorithm converges to the inverse because dynamically, it inverts the eigenvalues of $\mathbf{A}$. This also allows us to observe that the convergence time of the algorithm is dependent on the decay parameter $\alpha$ and the minimum eigenvalue $\lambda_\text{min}$ of $\mathbf{A}$, in addition to the time required to relax from the initial transient state.

\subsection{Moore-Penrose pseudo-inverse}
%\textcolor{red}{change this}
It is known ( see Showalter \cite{Showalter1967}) that the integral representation of the Moore-Penrose pseudoinverse is
\begin{equation}
    \mathbf{A}^\dagger = \int_0^\infty e^{(-\mathbf{A}^*\mathbf{A})t}\mathbf{A}^*\,dt
\end{equation}
Then by knowledge of the exact solution in (10) and using identical steps, we can see that to compute the pseudoinverse, the flow equation for the pseudoinverse should take the form
\begin{equation}
    \dot{\mathbf{R}} = -\alpha(\mathbf{A}^*\mathbf{A})\mathbf{R}+ \alpha  \mathbf{A}^*
\end{equation}
after discarding the $\mathbf{B}$ and $\mathbf{C}$ terms, meaning that the initial form of the forcing $\mathbf{V}$ should be
\begin{equation}
    \mathbf{V} = \alpha(-\mathbf{A}^*\mathbf{A} + B)\mathbf{R} - \mathbf{C} +  \alpha \mathbf{A}^*
\end{equation}
Solving for eqn. (\ref{eqapp:eq1}) with this forcing, we find the steady-state solution is
\begin{equation}
    \mathbf{R}(\infty)=\alpha \int_0^\infty e^{-\alpha(\mathbf{A}^*\mathbf{A})t} \mathbf{A}^*\, dt=\mathbf{A}^\dagger
\end{equation}
fitting Showalter's representation for $\mathbf{A}^\dagger$ \cite{Showalter1967}. Note that the coefficient $\alpha$ is necessary before the $\mathbf{A}^*\mathbf{A}$ term in the forcing in order to reduce the equation to the correct one.

\section{Stochastic analysis}\label{sec:stochan}
%\subsection{Perturbations to the Drazin inverse}
\subsection{Perturbations to the Drazin inverse}\label{app:perturbationdrazin}
We start with the vectorized equation
\begin{equation}
    \frac{d\vec r}{dt}=-\alpha \mathcal A \vec r+\alpha \vec e +\vec \xi(t)
\end{equation}
where
\begin{eqnarray}
\langle \xi_{ij}(t)\rangle&=&0\\
    \langle \xi_{ij}(t)\xi_{kl}(t^\prime)\rangle&=&2\sigma^2 \delta_{ik}\delta_{jl} \delta(t-t^\prime)
\end{eqnarray}
We have
\begin{eqnarray}
    \vec r(t) &=& e^{-\alpha t \mathcal A}\vec r(0) + \int_0^t e^{(t-\tau)(-\alpha \mathcal A)} (\alpha \vec e+\xi(\tau))\,d\tau
\end{eqnarray}
Let us make a change of variables inside the integral, and write $\tau=t-s$, $ds=-d\tau$. Then
\begin{eqnarray}
    \int_0^t e^{\alpha A(\tau-t)} d\tau=-\int_t^0 e^{-s\alpha A} ds=\int_0^t e^{-s\alpha A} ds.
\end{eqnarray}
\begin{eqnarray}
    \vec r(t) &=& e^{-\alpha t \mathcal A}\vec r(0) + \int_0^t e^{-s\alpha \mathcal A} (\alpha \vec e+\xi(t-s))\,ds
\end{eqnarray}
Now, we have
\begin{eqnarray}
\langle \vec r(t)\rangle=e^{-\alpha t \mathcal A}\vec r(0) + \int_0^t e^{-s\alpha \mathcal A} (\alpha \vec e)\,ds
\end{eqnarray}
and
\begin{eqnarray}
    \langle  r(t)\rangle &=& e^{-\alpha t \mathcal A}\vec r(0) + \alpha\int_0^t e^{-s\alpha \mathcal A}  \vec e\,ds
\end{eqnarray}

Now we use the fact that $\mathcal A=I_N\otimes A$. We have 
$e^{-\alpha t \mathcal A}=I_N\otimes e^{-t\alpha A}$. As a result, $\text{mat}(e^{-\alpha t \mathcal A}\vec r)=e^{-t\alpha A} R$. Then
\begin{eqnarray}
    \mathbf{R}(t) &=& e^{-\alpha t \mathbf{A}}\mathbf{R}(0) + \int_0^t e^{-s \alpha  \mathbf{A}}  (\alpha \mathbf{I} +\xi(t-s))\,ds\label{eq:mean1}
\end{eqnarray}

The solution is written then as
\begin{eqnarray}
    \langle  \mathbf{R}(t)\rangle &=& e^{-\alpha t  \mathbf{A}} \mathbf{R}(0) + \alpha\int_0^t e^{-\alpha \mathbf{A} s} ds
\end{eqnarray}

Now, let us add and subtract
\begin{eqnarray}
    \mathbf{R}(\infty)=\alpha \int_0^{\infty} e^{-\alpha \mathbf{A} s} ds= \int_0^{\infty} e^{-\mathbf{A} s} ds=\mathbf{A}^{D}
\end{eqnarray}
is the definition of the Drazin inverse of the matrix $\mathbf{A}$. If the matrix $\mathbf{A}$ is invertible, then $\mathbf{A}^D=\mathbf{A}^{-1}$.
We have
\begin{eqnarray}
    \mathbf{R}(\infty)-\langle  \mathbf{R}(t)\rangle &=& \alpha\int_{t}^\infty e^{-\alpha \mathbf{A} s} ds-e^{-\alpha t  \mathbf{A}}\vec R(0)
\end{eqnarray}
Let us now look at the fluctuations. We introduce the anticommutator $\{\mathbf{A},\mathbf{B}\}=\mathbf{A}\mathbf{B}+\mathbf{B}\mathbf{A}$.

We have, ignoring terms that are linear in $\xi$,
\begin{eqnarray}
\langle \mathbf{R}(t)\mathbf{R}(t^\prime)\rangle&=& \{e^{-\alpha t  \mathbf{A}} \mathbf{R}(0),e^{-\alpha t^\prime  \mathbf{A}} \mathbf{R}(0) \}  \\
&&+\alpha \big(e^{-\alpha t \mathbf{A}} \mathbf{R}(0)\int_0^{t^\prime} e^{-s^\prime \mathbf{A}} ds^\prime\nonumber \\
&+&e^{-\alpha t^\prime \mathbf{A}} \mathbf{R}(0)\int_0^{t} e^{-s \mathbf{A}} ds\big)\\
&+&\int_0^t\int_0^{t^\prime} ds ds^\prime \langle \{e^{-s \alpha  \mathbf{A}}  (\alpha I+\xi(t-s)),e^{-s^\prime \alpha \mathbf{A}}  (\alpha \mathbf{I}+\xi(t^\prime-s^\prime))\}\rangle\nonumber 
\end{eqnarray}
Now note that
\begin{eqnarray}
    &&\ \langle \{e^{-s \alpha \mathbf{A}}  (\alpha \mathbf{I}+\xi(t-s)),e^{-s^\prime \alpha \mathbf{A}}  (\alpha \mathbf{I}+\xi(t^\prime-s^\prime))\}\rangle=\alpha^2 e^{-(s+s^\prime) \alpha \mathbf{A}}\\
    &&+\langle \{e^{-s \alpha \mathbf{A}}  \xi(t-s),e^{-s^\prime \alpha \mathbf{A}}  \xi(t^\prime-s^\prime)\}\rangle
\end{eqnarray}
Since the integral is symmetric, we can simply focus on
\begin{eqnarray}
    \mathbf{M}=2 \langle e^{-s \alpha \mathbf{A}}  \xi(t-s) e^{-s^\prime \alpha \mathbf{A}}  \xi(t^\prime-s^\prime)\rangle
\end{eqnarray}
Let us make the indices explicit (we call the element $ij$ of the matrix $e^{M}$, $e_{ij}^M$ below):
\begin{eqnarray}
    \langle M_{ij}\rangle &=&2 \langle \sum_{klm} e^{-s \alpha  A}_{ik}  \xi_{kl}(t-s) e^{-s^\prime \alpha A}_{lm}  \xi_{mj}(t^\prime-s^\prime)\rangle\\
    &=&2 \sum_{klm} e^{-s \alpha A}_{ik}   e^{-s^\prime \alpha A}_{lm}  \langle \xi_{kl}(t-s) \xi_{mj}(t^\prime-s^\prime)\rangle\\
    &=&4\sigma^2 \sum_{klm} e^{-s \alpha A}_{ik}   e^{-s^\prime \alpha A}_{lm}  \delta_{km}\delta_{lj}\delta(t-t^\prime+s-s^\prime) \\
    &=&4\sigma^2 \sum_{k} e^{-s \alpha A}_{ik}   e^{-s^\prime \alpha  A^t}_{kj}  \delta(t-t^\prime+s-s^\prime)
\end{eqnarray}
Thus
\begin{eqnarray}
    \langle \mathbf{M}\rangle=4\sigma^2e^{-s \alpha \mathbf{A}} e^{-s^\prime \alpha \mathbf{A}^t}\delta(t-t^\prime+s-s^\prime)
\end{eqnarray}
where $A^t$ is the transpose. Thus
\begin{eqnarray}
    \langle \mathbf{R}(t)\mathbf{R}(t^\prime)\rangle&=&\{e^{-\alpha t  A} \mathbf{R}(0),e^{-\alpha t^\prime  \mathbf{A}} \mathbf{R}(0) \}\nonumber \\
&&+\alpha \big(e^{-\alpha t \mathbf{A}} \mathbf{R}(0)\int_0^{t^\prime} e^{-s^\prime \mathbf{A}} ds^\prime+e^{-\alpha t^\prime \mathbf{A}} \mathbf{R}(0)\int_0^{t} e^{-s \mathbf{A}} ds\big)\nonumber \\
   && +\alpha^2 \int_0^t \int_0^t dsds^\prime e^{-(s+s^\prime)\mathbf{A}}\\
    &&\ +4\sigma^2\int_0^tds \int_0^{t^\prime}d s^\prime\ e^{-s \alpha A} e^{-s^\prime \alpha \mathbf{A}^t}\delta(t-t^\prime+s-s^\prime)
\end{eqnarray}
Now, we note that the first tree lines arise from $\langle \mathbf{R}(t)\rangle\langle \mathbf{R}(t^\prime)\rangle$.
Thus
\begin{eqnarray}
    \langle \mathbf{R}(t)\mathbf{R}(t^\prime)\rangle-\langle \mathbf{R}(t)\rangle\langle \mathbf{R}(t^\prime)\rangle=4\sigma^2\int_0^tds \int_0^{t^\prime}d s^\prime\ e^{-s \alpha \mathbf{A}} e^{-s^\prime \alpha \mathbf{A}^t}\delta(t-t^\prime+s-s^\prime)
\end{eqnarray}
In the case $t=t^\prime$ we have
\begin{eqnarray}
    \Delta \mathbf{R}(t)^2&=&4\sigma^2\int_0^tds\  e^{-s \alpha \mathbf{A}} e^{-s\alpha \mathbf{A}^t}
\end{eqnarray}
If the matrix $\mathbf{A}$ is symmetric, it reduces to
\begin{eqnarray}
    \Delta \mathbf{R}(t)^2&=&4\sigma^2\int_0^tds\  e^{-2s \alpha \mathbf{A}} 
\end{eqnarray}
Thus, the fluctuation over the mean for uncorrelated white noise at large times is
\begin{eqnarray}
    \Delta \mathbf{R}(\infty)^2&=& \frac{2\sigma^2}{\alpha} \langle \mathbf{R}(\infty)\rangle
\end{eqnarray}

\subsection{Case of Moore-Penrose inverse}
We now consider the perturbation of the dynamics leading to the Moore-Penrose equation, e.g. 
\begin{eqnarray}
\frac{d\mathbf{R}}{dt}&=&-\alpha \mathbf{A}^t \mathbf{A} \mathbf{R}+\alpha \mathbf{A}^t+\xi(t).
\end{eqnarray}
The solution is given by a simple generalization of eqn. (\ref{eq:mean1}):
\begin{eqnarray}
    \mathbf{R}(t) &=& e^{-\alpha t \mathbf{A}^t \mathbf{A}}\mathbf{R}(0) + \int_0^t e^{-s \alpha \mathbf{A}^t \mathbf{A}}  (\alpha \mathbf{A}^t+\xi(t-s))\,ds
\end{eqnarray}
We are interested in the average
\begin{eqnarray}
    \langle \mathbf{R}(t)\mathbf{R}(t^\prime)\rangle &=& \langle (e^{-\alpha t \mathbf{A}^t \mathbf{A}}\mathbf{R}(0) + \int_0^t e^{-s \alpha \mathbf{A}^t \mathbf{A}}  (\alpha \mathbf{A}^t+\xi(t-s))\,ds)\\
    && \cdot(e^{-\alpha t^\prime \mathbf{A}^t \mathbf{A}}\mathbf{R}(0) + \int_0^{t^\prime} e^{-s \alpha \mathbf{A}^t \mathbf{A}}  (\alpha \mathbf{A}^t+\xi(t^\prime-s))\,ds)\rangle
\end{eqnarray}
Note that we have
\begin{eqnarray}
    \langle \mathbf{R}(t)\rangle\langle \mathbf{R}(t^\prime)\rangle &=& \ (e^{-\alpha t \mathbf{A}^t \mathbf{A}}R(0) + \alpha\int_0^t e^{-s \alpha \mathbf{A}^t \mathbf{A}}   \mathbf{A}^t\,ds)\\
    && \cdot(e^{-\alpha t^\prime \mathbf{A}^t \mathbf{A}}\mathbf{R}(0) + \alpha\int_0^{t^\prime} e^{-s \alpha \mathbf{A}^t \mathbf{A}}   \mathbf{A}^t\,ds)
\end{eqnarray}
The result here is a little bit simpler as $A^t A$ is symmetric, and thus
\begin{eqnarray}
    \langle \mathbf{R}(t)\mathbf{R}(t^\prime)\rangle-\langle \mathbf{R}(t)\rangle\langle \mathbf{R}(t^\prime)\rangle&=& 4\sigma^2\int_0^t \int_0^{t^\prime}ds ds^\prime e^{-\alpha (s+s^\prime)\mathbf{A}^t \mathbf{A}}\delta(t-t^\prime+s-s^\prime)\nonumber
\end{eqnarray}
For $t=t^\prime$ we have
\begin{eqnarray}
    \Delta \mathbf{R}(t)^2= 4 \sigma^2 \int_0^t e^{-2 s \alpha \mathbf{A}^t A}ds=\frac{2 \sigma^2}{\alpha}\int_0^t e^{- s  \mathbf{A}^t \mathbf{A}}ds
\end{eqnarray}
which is the formula reported in the main text.
Thus, at long times the fluctuations are proportional to the Drazin inverse of $\mathbf{A}^t \mathbf{A}$. Since $\mathbf{A}^t \mathbf{A}$ is invertible, then this is just the inverse of $\mathbf{A}^t \mathbf{A}$.

\section{Stability analysis}
We now discuss the stability of the algorithm if volatile devices are present.

In the analysis of dynamical systems, the poles of the transfer function play a critical role in determining system stability. The transfer function, $ H(s) $, obtained through the Laplace transform of the system's differential equations, is expressed as a ratio of polynomials in the complex frequency variable $ s $. The poles are the roots of the denominator polynomial, representing the values of $ s $ for which $ H(s) $ becomes unbounded. Mathematically, if $ H(s) = \frac{N(s)}{D(s)} $, the poles are the solutions to $ D(s) = 0 $. The location of these poles in the complex plane directly influences the stability: if all poles lie in the left half of the complex plane (i.e., have negative real parts), the system is stable as perturbations decay over time. Conversely, poles with positive real parts indicate instability, causing perturbations to grow. Poles on the imaginary axis suggest marginal stability, where perturbations neither decay nor grow but persist indefinitely. Thus, the analysis of the poles provides crucial insights into the dynamic behavior and stability characteristics of the system. 

\subsection{Realistic function on the right-hand side}
\begin{figure}
    \centering
    \includegraphics[width=0.7\linewidth]{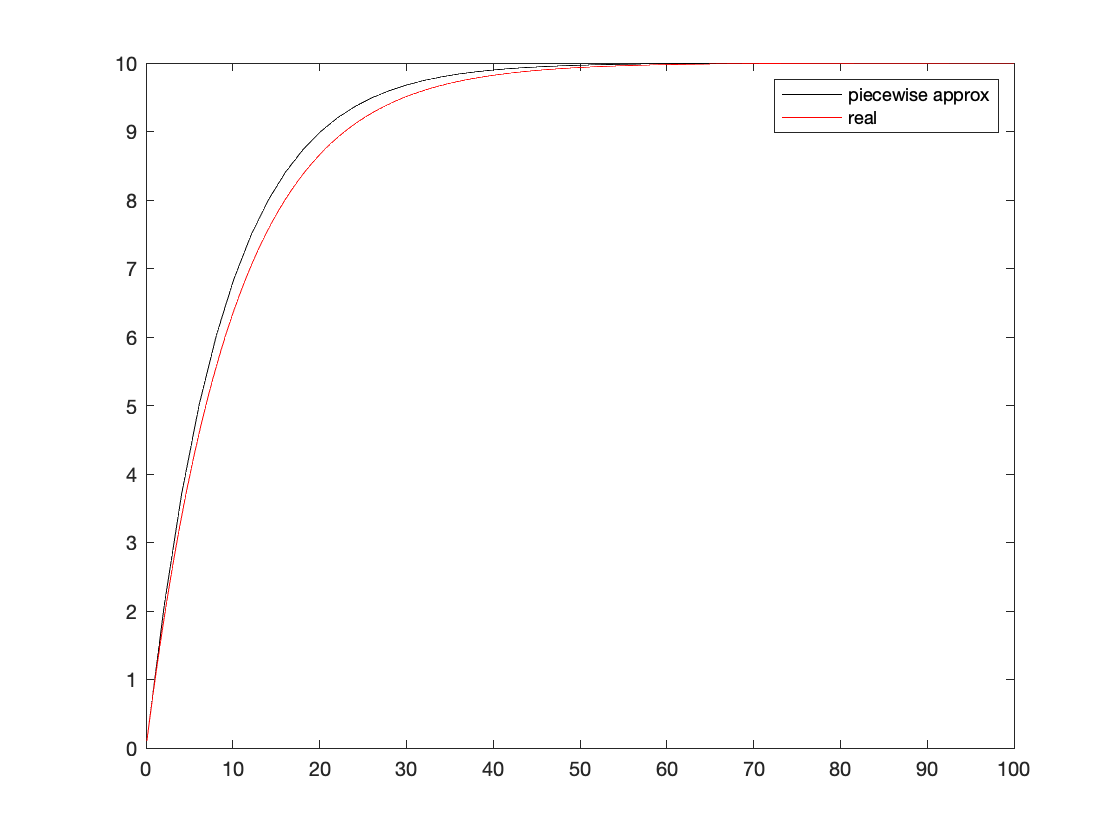}
    \caption{Evolution for a delayed controller with a piecewise approximate versus instantaneous.}
    \label{fig:checkconv}
\end{figure}

Instead of $x(t-\tau)$, the function that should be applied is of the form, for a certain $\tau$
\begin{eqnarray}
    f_\tau(t-\tau)=\sum_{k=0}^{\infty} f((k-1)\tau)\big(\theta(t-(k-1)\tau)-\theta(t-k\tau)\big)
\end{eqnarray}
In the limit $\tau\rightarrow 0$, $\lim_{\tau\rightarrow 0}f_\tau(t-\tau)=f(t)$ pointwise.
The Laplace transform is
\begin{eqnarray}
    f_\tau(s)&=&\mathcal L(f_\tau)=\sum_{k=0}^{\infty} f((k-1)\tau)\frac{\big(e^{-(k-1) \tau s}-e^{-k \tau s}\big)}{s} \nonumber \\
    &=&\sum_{k=0}^{\infty} f((k-1)\tau)e^{-k \tau s}\frac{\big(e^{\tau s}-1\big)}{s}
\end{eqnarray}
If the delay $\tau$ is sufficiently small, we can identify $ k\tau=t$ and $\tau =dt$, and it is easy to see that
\begin{eqnarray}
    f_\tau(s)=\mathcal L(f_\tau)=\int_0^\infty dt\ f(t-\tau) e^{-t s}+O(\tau)=e^{-\tau s}\mathcal L(f)+O(\tau).
\end{eqnarray}
We simulated eqn. (\ref{eq:oned}) with a small delay and a piecewise approximation for $x(t-\tau)$ as a controller would do. We see in Fig. \ref{fig:checkconv} that the asymptotic point is the same.

\subsection{Laplace transform for scalar linear delayed ODE}\label{app:delayana} 
\begin{figure}
    \centering
    \includegraphics[width=0.5\linewidth]{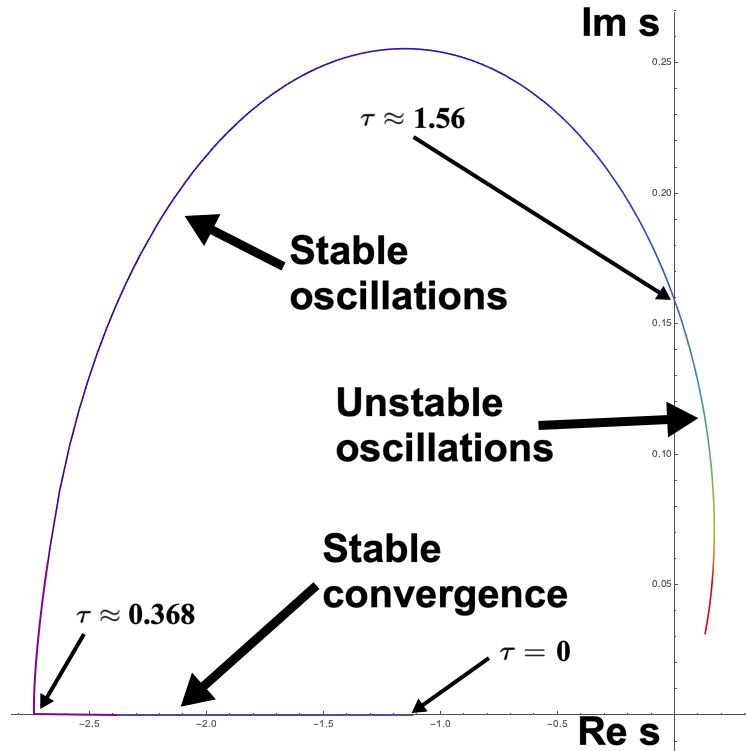}
    \caption{Phase oscillations and instability for the simple equation of eq. (\ref{eq:oned}).}
    \label{fig:phase}
\end{figure}

Let us look at the delayed scalar differential equation:
\begin{eqnarray}
    \frac{dx}{dt}=-\alpha a x(t-\tau)+b.\label{eq:oned}
\end{eqnarray}
Using the Laplace transform we have
\begin{eqnarray}
    sx(s)-x_0=-\alpha a e^{-\tau s} x(s)+\frac{b}{s}
\end{eqnarray}
from which we get the transfer function
\begin{eqnarray}
    x(s)=\frac{s x_0+b}{s(s+\alpha a e^{-\tau s})}
\end{eqnarray}
The two poles of this transfer function are in $s_1=0$ and the solution of
\begin{eqnarray}
    s+\alpha a e^{-\tau s}=0\label{eq:scalarpoles}
\end{eqnarray}
whose solution is given by
\begin{eqnarray}
    s_2=\frac{W(-a \alpha  \tau )}{\tau }
\end{eqnarray}
where $W$ is the product-log function. In the limit $\tau\rightarrow 0$, $s_2\rightarrow-a \alpha$ which is real.
However, as $\tau$ increases the function can develop into the complex plane.
If a pole arises in the complex plane in the transfer function, then oscillations occur. Using this method, we can calculate the phase diagram in terms of the delay. Let us set $a\alpha=1$, and measure $\tau$ in units of $a\alpha$. This can be seen in Fig. \ref{fig:phase}.

\subsection{Poles of the matricial transfer function}
\begin{figure}
    \centering
    \includegraphics[width=0.5\linewidth]{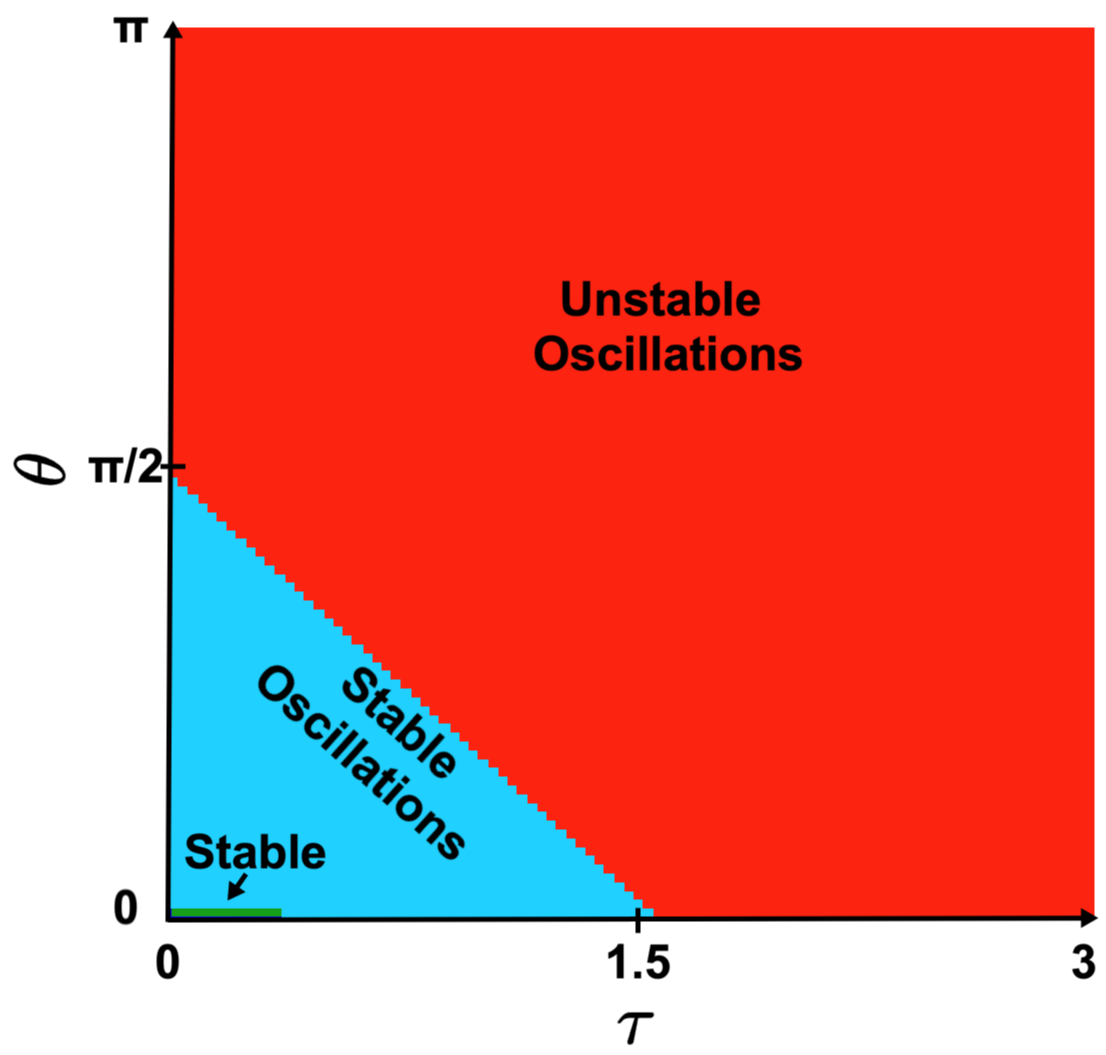}
    \caption{Phase diagram of the stability of the delayed ODE of eqn. (\ref{eq:dode}) with $\xi_P=\xi_O=0$. Specifically, we analyze the angular dependence of the pole in the complex plane and identify the stability. The plot is symmetric on the $[0,2\pi]$ interval. As we can see, for short delays the system is stable, while at slightly longer delays the system is marginally stable. At much longer delays there is an onset of instability.}
    \label{fig:complex}
\end{figure}

Let us apply the Laplace transfer method to the matricial equation with a small delay as an approximate method.
We have
\begin{equation}
    s\mathbf{R}(s)-\mathbf{R}(0)=-\alpha \mathbf{A} \mathbf{R}(s) e^{-\tau s}+ \frac{\alpha}{s} \mathbf{I}
\end{equation}
and thus 
\begin{eqnarray}
    \mathbf{R}(s)&=&(s \mathbf{I}+\alpha \mathbf{A} e^{-\tau s})^{-1} (\mathbf{R}(0)+\frac{\alpha }{s} \mathbf{I})\\
    &=&\frac{1}{s}(s \mathbf{I}+\alpha \mathbf{A} e^{-\tau s})^{-1} (s\mathbf{R}(0)+\alpha  \mathbf{I}),
\end{eqnarray}
where the transfer function is now a matrix. We then need to look at the poles of the inverse matrix. These can be found via
\begin{eqnarray}
    \text{det}(s \mathbf{I}+\alpha \mathbf{A}e^{-\tau s})=0.
\end{eqnarray}
This is the matricial equivalent of eqn. (\ref{eq:scalarpoles}). We recall that the eigenvalues of a matrix $A$ are the solution of $\text{det}(\lambda \mathbf{I}-\mathbf{A})=0$.
Let us call $\lambda=-\frac{s e^{\tau s}}{\alpha}$ for $s\neq 0$. 
We can find the poles by looking at the eigenvalues of the matrix $A$, and identifying
\begin{eqnarray}
    -\alpha \lambda_i= s e^{\tau s}.
\end{eqnarray}
We thus see that the problem of the stability of the system is equivalent to the scalar problem, e.g.
\begin{eqnarray}
    s_i=\frac{W(-\alpha \lambda_i \tau)}{\tau},
\end{eqnarray}
with the caveat that $\lambda_i$ now can be complex.
If the matrix $A$ is normal and eigenvalues are real, then the analysis of the previous section is valid. If instead, the eigenvalues are complex the analysis becomes more complicated. Let us choose $|\lambda_i|=\rho_i$. Fixing $|\alpha \rho_i|=1$, the graph of the poles as a function of $\tau$ and $\theta$ is shown in Fig. \ref{fig:complex} where we wrote $\lambda_i=\rho_i e^{i\theta_i}$.

\section{Crossbar implementation}
\subsection{Calibration}

The computed forcing requires precise knowledge of the decay and steady-state parameters $\alpha$ and $C$. It is therefore desirable to be able to quickly measure these properties, preferably using a protocol that is simple and requires minimal computational complexity such that it may be repeated as necessary. The following protocol requires only two memristor reads, basic arithmetic, a simple constant forcing, and takes time on the order of $\Theta(2\alpha)$.

Suppose we allow an unforced memristor to relax fully to its steady-state resistance $R(\infty) = \frac{C}{\alpha}$. Let us call this value $R_1$.

Now suppose we force the memristor with a voltage $V = R_1$. It is easy to see from (\ref{eqapp:eq1}) that the steady-state value $R_2$ of the memristor under this forcing is
\begin{equation}
    R_2 = \frac{C + \frac{C}{\alpha}}{\alpha} = \frac{C(1 + \alpha)}{\alpha^2}
\end{equation}
We now note that
\begin{align}
    \frac{R_1}{R_2} &= \frac{C}{\alpha}\cdot\frac{\alpha^2}{C(1+\alpha)} \\
    &= \frac{\alpha}{1 + \alpha}
\end{align}
and thus clearly $\alpha = \left(\frac{R_2}{R_1} - 1\right)^{-1} \rightarrow C = \alpha R_1$. We see that the speed of this scheme is primarily determined by the time it takes the memristors to relax to steady state. 

On-chip, arithmetic operations can be implemented by ADCs and common integrated circuits. The scheme can also be performed elementwise for maximum accuracy.

\subsection{Wake up signal for volatile devices}
\begin{figure}
    \includegraphics[width=0.49\textwidth]{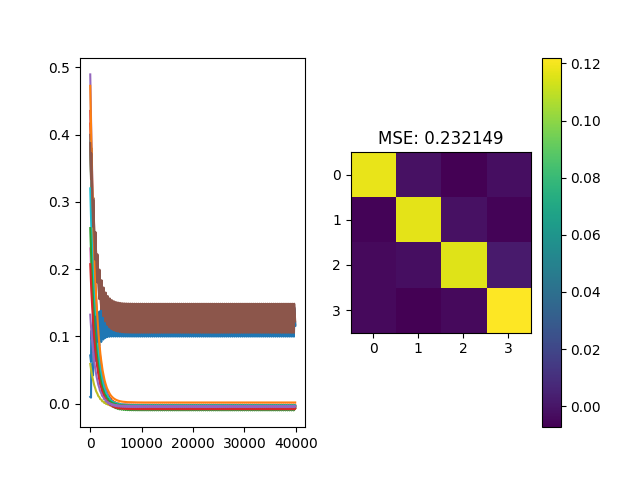}
    \includegraphics[width=0.49\textwidth]{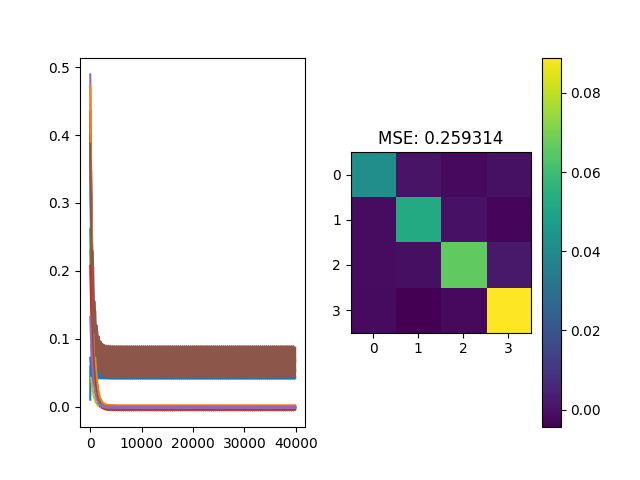}
    \caption{Simulation of matrix inversion done na{\"i}vely on a crossbar containing isolated volatile crossbar elements (versus the nonvolatile crossbars considered in the main text). Because each subset of elements has zero bias while a different subset is being written, it decays freely and the matrix inverse fails to converge. \textbf{Left.} Inversion results when entire diagonals of elements are written simultaneously (i.e. $O(N)$). \textbf{Right.} Inversion results when each element is written sequentially in $O(N^2)$.}
    \label{fig:dde_vol_decay}
\end{figure}

\begin{figure}
    \includegraphics[width=0.8\textwidth]{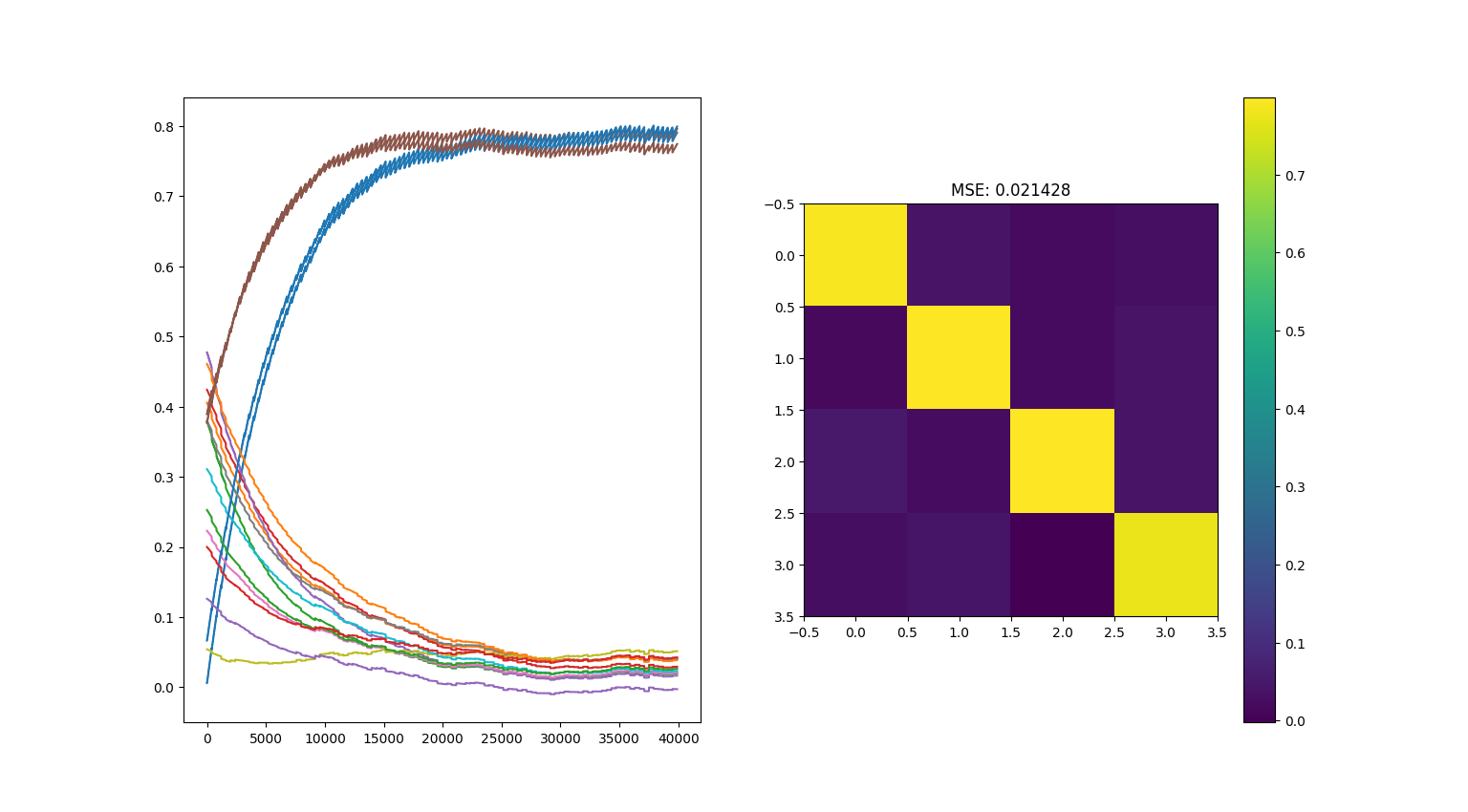}
    \caption{When each volatile element is forced with a higher voltage to compensate for decays during period of zero bias, the crossbar state again converges to the desired inverse $\mathbf{A}^{-1}$.}
    \label{fig:conv_ideal_delayed}
\end{figure}

We now consider the implementation of this algorithm in a physical crossbar array, in which volatile memristors share input buses. It is clear that only a subset of memristors may be forced simultaneously to achieve high-rank updates to the state; moreover, a finite amount of time must be spent on each subset to write their memory parameters adequately. There is then a clear obstacle to convergence in the event that the system comprises memristors with high decay parameter $\alpha$ and relatively high subset writing time $\tau + \theta$, where $\tau$ represents the switching time of a transistor gating individual memristors and $\theta$ represents the width of a square wave pulse applied to the memristor. In this situation convergence is significantly hampered by the fact that the memory values of one subset will decay while another is being driven. Indeed, we observe this to be the case in simulation curves shown in Figure \ref{fig:dde_vol_decay}; the decay of inactive subsets significantly outpaces the contributions from the forcing in (\ref{eqapp:eq1}).

It is thus necessary to further modify the forcing to account for the decays during inactivity; the general idea is that each subset of memristors must be driven higher than in (\ref{eqapp:eq1}) to compensate for decays while the other subsets are being written. We refer to this modification as a wake up signal. %We refer to this modification as \textit{backwards reverse compensatory mitigatory fixative wakeup preventative preemptive prophylactic forcing}.

We begin by finding the state of an unforced memristor relaxing from state $R(0)$ at some time $\tau$ according to (\ref{eq:drazin_tr}), setting $V=0$ to find
\begin{align}
    R(\tau) &= e^{-\alpha\tau}R(0) + C\int_0^{\tau} e^{-\alpha(\tau - t)}\, dt \label{eq:mem_unforced} \\
    &= e^{-\alpha\tau}R(0) + Ce^{-\alpha\tau}\left(\frac{1}{\alpha}e^{\alpha t}\Bigg|^\tau_0\right) \\
    &= e^{-\alpha\tau}R(0) + \frac{Ce^{-\alpha\tau}}{\alpha}(e^{\alpha\tau} - 1) \\
    &= e^{-\alpha\tau}\left[\frac{\alpha R(0)+C(e^{\alpha\tau}-1)}{\alpha}\right]
\end{align}
At the end of constant forcing $V$ over the interval $[T_1, T_2]$, the final resistance is thus
\begin{equation}
    R_\text{force}(T_2) = e^{-\alpha\tau}\left[\frac{\alpha R(T_1)+(C+V)(e^{\alpha\tau}-1)}{\alpha}\right]
    \label{eq:force_tau}
\end{equation}
and it is easy to see that given an initial resistance $R(T_1)$, the constant voltage required to drive the memristor to a target value $R_\text{force}(T_2)$ is
\begin{equation}
    V(R_\text{force}(T_2)) = \left(\frac{\alpha R_\text{force}(T_2)}{e^{-\alpha T}} - \alpha R(T_1)\right)\frac{1}{e^{\alpha T}-1} - C
    \label{eq:modtarget}
\end{equation}
where $T = T_2 - T_1$. Finally we observe from (\ref{eq:mem_unforced}) that for (\ref{eq:force_tau})
\begin{equation}
    \lim_{\tau \rightarrow \infty} R_\text{force}(\tau) = (C+V)\int_0^{\infty} e^{-\alpha t}\, dt = \frac{C+V}{\alpha}
\end{equation}
and we may progress to modifying the forcing.

We now consider a case in which the restricted writing subsets of a crossbar matrix $R$ are its columns, such that a complete write of the crossbar takes time $N(\tau+\theta)$. Each column thus experiences a delay of duration $(N-1)(\tau+\theta)$ before it gets written again.

Suppose that at a single write step, we have already quickly read the crossbar and computed the desired forcing, e.g. according to (\ref{eqapp:eq1}). Let this forcing be $V^{\text{(ideal)}}\in \mathbb{R}^{N \times N}$.

We now ask what resistances the array $R$ would be driven to if $V^{\text{(ideal)}}$ were applied infinitely. Using (68), we find that under this forcing $R^\text{(ideal)}_{ij}(\infty) = \frac{C + V_{ij}}{\alpha}$. 

We wish to find $R^\text{(pre)}(0)$ such that $R((N-1)(\tau+\theta)) = R^\text{(ideal)}(\infty)$ before the next write. We note that this value is just (\ref{eq:force_tau}) with a negative time input, i.e. evaluating $R(-(N-1)(\tau+\theta))$ with $R(0) = R^\text{(ideal)}(\infty)$. We therefore drive each element $R_{ij}$ of $R$ to be
\begin{align}
    R^\text{(pre)}_{ij} &= e^{\alpha(N-1)(\tau+\theta)}\left[\frac{\alpha R_{ij}^\text{(ideal)}(\infty)+C(e^{-\alpha(N-1)(\tau+\theta)}-1)}{\alpha}\right] \\
    &= e^{\alpha(N-1)(\tau+\theta)}\left[\frac{\alpha\left(\frac{C+V^\text{(ideal)}_{ij}}{\alpha}\right)+C(e^{-\alpha(N-1)(\tau+\theta)}-1)}{\alpha}\right] \\
    &= e^{\alpha(N-1)(\tau+\theta)}\left[\frac{C+V^\text{(ideal)}_{ij}+C(e^{-\alpha(N-1)(\tau+\theta)}-1)}{\alpha}\right]
\end{align}
which yields our modified forcing
\begin{equation}
    V_{ij}^\text{(pre)}(R_{ij}^\text{(pre)}) = \left(\frac{aR_{ij}^\text{(pre)}}{e^{-\alpha T}} - \alpha R_{ij}\right)\frac{1}{e^{\alpha T}-1} - C
\end{equation}
which converges to the inverse even in the presence of subset decay (Figure \ref{fig:conv_ideal_delayed.png}).

%\subsection{Issues with convergence}
%\input{secs/Convergence}

\subsection{Gating requirements for analog switches}
The algorithm assumes that memristors are perfectly isolated from each other during reading and writing. In reality, the analog switches used to achieve 1T1R gating have a finitely high ``off" resistance. This introduces an error with weak dependence on system size $N$, as in fact nonzero current flows through the entire system during all operations. While this current is extremely small, it becomes relevant in the context of highly nonlinear operations like matrix inversion. This problem is also commonly referred to as the ``sneak path" problem in crossbars.

We can investigate the required value of $R_\text{off}$ for the gating switches by varying between $\qty{1e6}{\ohm}$ and $\qty{1e9}{\ohm}$ and observing the final dependence on system size.

\subsection{Atomic operations}\label{app:atomop}
We define an atomic operation as an update of the values output by the $N$ voltage sources in the crossbar, and an update of the biases on the $N^2$ switches gating each element. A matrix-vector product $Gv$ is thus $O(1)$ because it requires a single update of the $N$ sources (to be the elements of $v$) and a single update of the $N^2$ transistors (biasing all gates). Likewise, read and write operations are $O(N)$ because they require $2N - 1$ distinct updates of the transistor states (and of the voltage sources in the latter case).

In general, a single ``iteration" comprises a read of the matrix state $R$, a computation of the matrix product $AR$, and an update employing this product (e.g. $-\alpha AR + \alpha I$). All these operations are in $O(N)$ by the definition above, so a single iteration is in $O(N)$.

\subsection{Mixed-Sign Computation}\label{sec:mixed_sign}
\begin{figure}
    \includegraphics[width=\textwidth]{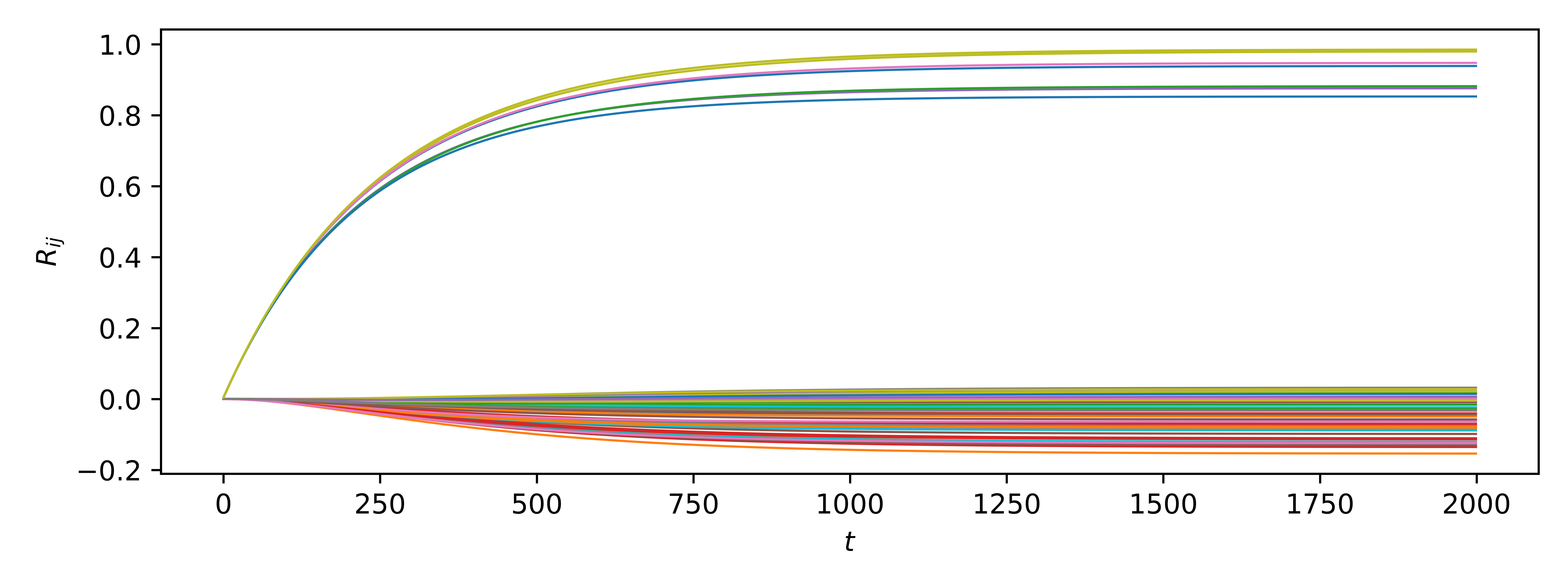}
    \caption{When terms of $A$ are of mixed sign, the terms $R_{ij}$ converge monotonically to their respective destinations from 0.}
    \label{fig:monotonic_inverse}
\end{figure}

In general, $A$ will have both positive and negative terms, while a given crossbar can only store terms of positive sign. It is therefore necessary to employ two crossbars for general computation, in which the matrix $A$ is decomposed into postive and negative-only matrices such that $A = A^+ + A^-$, as suggested in \cite{Sun2019}. Suppose we can only store positive-valued matrix entries. Then to compute a matrix-vector product we have $Av = (A^+ + A^-)v = A^+v - |A^-|v$, where the final expression follows because $A^-_{ij} \leq 0$. This practice is common for any form of single-sign storage medium.

What is more interesting is that this approach can be applied immediately to this iterative algorithm, provided that both $R^+$ and $R^-$ are initialized to $\mathbf{0}$. When initialized in this way, the algorithm evolves such that $R_{ij}$ either increase or decrease monotonically (Figure \ref{fig:monotonic_inverse}), such that positive terms only experience positive gradients and negative terms likewise. Then if $R^+$ and $R^-$ are crossbars containing memristors of opposite polarity, they will naturally separate to the correct values.

\subsection{Parameter learning}\label{sec:paramlearnapp}
We will use variational inference as a method to learn a Gaussian distribution. We assume that we are given a set of continuous data $\vec x_i\in \mathbb{R}^N$. This means that there is a method (experiments, a sampler, etc) from which, given a distribution $p(\vec x)$, we obtain a set of points $\vec x_i$ sampled from $p(\vec x)$. 
Our assumption in this section in this is that $p(\vec x)$ is Gaussian, e.g.
\begin{eqnarray}
    p_\mathbf{A}(\vec x)&=&\frac{1}{Z(A)} e^{-\frac{1}{2} \vec x^t \mathbf{A} \vec x}\\
    Z(A)&=&\int dx^N e^{-\frac{1}{2} \vec x^t \mathbf{A} \vec x}.
\end{eqnarray}
The question is how can we learn the parameters of this model, e.g. $A_{ij}$, which is called parameter estimation.

The method we will discuss here is variational inference. Let us look at the \index{key}{KL divergence}, which will be our function to minimize to obtain the parameters:
\begin{eqnarray}
\mathcal L(\mathbf{A})=D_{KL}(p\|p_\mathbf{A})=\int dx^N\ p(\vec x) \log \frac{p(x)}{p_\mathbf{A}(x)}.   
\end{eqnarray}
We know that $\mathcal L(\mathbf{A})=0$ if and only if $p(x)=p_\mathbf{A}(x)$. However, we already see a few problems here. First, we do not know $p(x)$. However, we can estimate the average from our samples. 
First, note that the term
\begin{eqnarray}
    p(x) \log p(x)
\end{eqnarray}
does not depend on $\mathbf{A}$, and thus from the point of view of minimization of $\mathcal L$ it does not contribute, as it is just a constant. Thus, we are left with
\begin{eqnarray}
\mathcal L(\mathbf{A})=D_{KL}(p\|p_\mathbf{A})=-\langle \log p_\mathbf{A}(x)\rangle_{p}+constant
\end{eqnarray}
where the constant is intended as constant for $A$.
Second, we note that
\begin{eqnarray}
    \langle O(x)\rangle_{p(x)}\approx \frac{1}{M} \sum_{i=1}^M O(\vec x_i)
\end{eqnarray}
where $M$ is the number of samples. From Chebyshev's inequality, we know that
\begin{eqnarray}
    P(|\frac{1}{N} \sum_i x_i-\mu|<\epsilon)\leq \frac{\sigma^2}{n \epsilon^2}
\end{eqnarray}
Thus, we can replace the analytical mean with an empirical one, given the assumption that the number of samples is large enough and the variance is not scaling with $M$.

We rewrite then
\begin{eqnarray}
\mathcal L_{emp}(\mathbf{A})&=& -\frac{1}{M}\sum_{i=1}^{M} \log e^{-\frac{1}{2} \vec x_i^t \mathbf{A} \vec x_i}+ \log Z(\mathbf{A})\\
&=&-\frac{1}{M}\sum_{ab} \sum_{i=1}^M \frac{1}{2} (x_i)_a A_{ab} (x_i)_b+\log Z(\mathbf{A}).
\end{eqnarray}
Since our distribution is Gaussian, we have
\begin{eqnarray}
    Z(\mathbf{A})&=&\frac{(2\pi)^N}{\sqrt{\text{det}(\mathbf{A})}}\\
     \log Z(\mathbf{A})&=&\log \sqrt{2\pi}-\frac{1}{2} \log \text{det}(\mathbf{A})
\end{eqnarray}
The question then arises of how to minimize this functional. 
We will use a variational method. We minimize step by step the empirical KL divergence by following the gradient, e.g.
\begin{eqnarray}
    \frac{\partial \mathcal L_{emp}(\mathbf{A}) }{\partial A_{ab}}=-\frac{1}{M}\sum_{i=1}^M (x_i)_a(x_i)_b+\partial_{A_{ab}} \log Z(\mathbf{A})
\end{eqnarray}
Let us write $1/M\sum_{i=1}^M (x_i)_a(x_i)_b=\langle x_a x_b\rangle_{emp}$. We note immediately that this is the empirical correlator. We now recall then that
\begin{eqnarray}
    \partial_{A_{ab}} \log Z(\mathbf{A})=(A)^{-1}_{ab}.
\end{eqnarray}
As a result, it is possible to write the following gradient
algorithm
\begin{eqnarray}
    \frac{d(A_t)_{ab}}{dt}=-\xi \frac{\partial \mathcal L_{emp}(\mathbf{A}) }{\partial A_{ab}}=\xi(\langle x_a x_b\rangle_{emp}-(A_t^{-1})_{ab})
    \label{eq:inversegaussiannooff}
\end{eqnarray}
where $\xi$ is a parameter sometimes called \textit{learning rate}.
This method is called \index{key}{variational inference} variational inference and we have seen here in action for the simplest example. Of course, this method works if $\langle x\rangle_{emp}=0$.

%\subsection{SPICE simulations and 1T1R %scheme} \label{sec:spice}
%\input{secs/SPICE}

% References

\end{document}